\documentclass[aps,nofootinbib,prd,superscriptaddress,twocolumn,10pt,preprintnumbers]{revtex4-1}

\usepackage{epsf,epsfig,subfigure,graphicx,amsmath,amssymb}
\usepackage{amsfonts}
\usepackage{latexsym}
\usepackage{color,xcolor}
\usepackage{accents}
\usepackage{geometry}
\usepackage{dhucs}
\usepackage{aas_macros}
\usepackage{hyperref}
\usepackage{verbatim}
\usepackage{slashed}
\usepackage{notes2bib}
\usepackage{tikz}
\usepackage{mathtools}
\usepackage{float}
\usepackage{soul} 

\def\bea{\begin{eqnarray}}
\def\eea{\end{eqnarray}}

\def\eV{\,{\rm eV}}

\def\bfx{{\bf x}}

\def\bfv{{\bf v}}

\def\eps{\epsilon}

\begin{document}
\preprint{DESY-22-200}

\title{Adiabatically compressed wave dark matter halo
\\and intermediate-mass ratio inspirals}

\author{Hyungjin Kim}
\email{hyungjin.kim@desy.de}
\affiliation{Deutsches Elektronen-Synchrotron DESY, Notkestr.85, 22607 Hamburg, Germany}

\author{Alessandro Lenoci}
\email{alessandro.lenoci@desy.de}
\affiliation{Deutsches Elektronen-Synchrotron DESY, Notkestr.85, 22607 Hamburg, Germany}

\author{Isak Stomberg}
\email{isak.stomberg@desy.de}
\affiliation{Deutsches Elektronen-Synchrotron DESY, Notkestr.85, 22607 Hamburg, Germany}

\author{Xiao Xue}
\email{xiao.xue@desy.de}
\affiliation{II. Institute of Theoretical Physics, Universit{\"a}t Hamburg, 22761 Hamburg, Germany}
\affiliation{Deutsches Elektronen-Synchrotron DESY, Notkestr.85, 22607 Hamburg, Germany}

\
\begin{abstract}
The adiabatic growth of a central massive black hole could compress the surrounding dark matter halo, leading to a steeper profile of the dark matter halo. 
This phenomenon is called {\it adiabatic compression}.
We investigate the adiabatic compression of wave dark matter -- a light bosonic dark matter candidate with its mass smaller than a few eV.
Using the adiabatic theorem, we show that the adiabatic compression leads to a much denser wave dark matter halo similar to the particle dark matter halo in the semiclassical limit.
The compressed wave halo differs from that of the particle halo near the center where the semiclassical approximation breaks down, and the central profile depends on dark matter and the central black hole mass as they determine whether the soliton and low angular momentum modes can survive over the astrophysical time scale without being absorbed by the black hole. 
Such a compressed profile has several astrophysical implications.
As one example, we study the gravitational waves from the inspiral between a central intermediate-mass black hole and a compact solar-mass object within the wave dark matter halo.
Due to the enhanced mass density, the compressed wave dark matter halo exerts stronger dynamical friction on the orbiting object, leading to the dephasing of the gravitational waves.
The pattern of dephasing is distinctive from that of inspirals in the particle dark matter halo because of the difference in density profile and because of the relatively suppressed dynamical friction force, originating from the wave nature of dark matter.
We investigate the prospects of future gravitational wave detectors, such as Laser Interferometer Space Antenna, and identify physics scenarios where the wave dark matter halo can be reconstructed from gravitational wave observations. 
\end{abstract}

\maketitle

\tableofcontents

\section{Introduction}
The cold dark matter paradigm successfully explains large-scale structures of the universe.
Observational results from several hundred-kiloparsec scales to megaparsec scales can be explained by introducing collisionless non-luminous matter, which takes roughly $30$\% of total energy density in the present universe. 
The evolution of dark matter on galactic scales becomes nonlinear at some point, and it eventually collapses to form dark matter halos, hosting galaxies in the universe.
Numerical simulations found a universal profile for the dark matter halo, which follows a broken power-law; $\rho \propto r^{-\gamma}$ with $\gamma \sim 3$ for the outer part and with $\gamma \sim 1$ for the inner part of the halo~\cite{1996ApJ...462..563N, 1997ApJ...490..493N}.\footnote{The inner slope of dark matter halo of observed galaxies might be shallower than $\gamma\sim1$. See Ref.~\cite{Salucci:2018hqu} for a recent review}

Much less is known for the structure of dark matter halo on sub-galactic scales.
Indeed, the inner part of the dark matter halo can be affected by the evolution of baryonic matter.
If a black hole at the center of the system has formed and grown adiabatically or a baryonic gas cloud sinks to the center of the system through dissipation, the inner profile of dark matter halo could be much steeper than its initial profile~\cite{1980ApJ...242.1232Y, 1986ApJ...301...27B, 1995ApJ...440..554Q, 1999AJ....117..744V}.
This phenomenon -- {\it adiabatic compression} -- leads to a dark matter halo with a characteristic cusp $\rho \propto r^{-\gamma_{\rm sp}}$ with $\gamma_{\rm sp} > \gamma$. 
This cuspy profile is often referred to as dark matter spike to avoid possible confusion with a usual NFW-like cuspy profile. 

Such a steeper dark matter profile provides interesting opportunities for new physics searches.
Since the dark matter density is greatly enhanced, it may enhance dark matter indirect signals~\cite{Gondolo:1999ef}.
Due to the enhanced mass density, celestial objects orbiting inside the adiabatically compressed halo experience larger dynamical friction.
Such dynamical friction could change the gravitational wave signals emitted from black hole binaries embedded in the halo~\cite{Eda:2013gg, Eda:2014kra, Kavanagh:2020cfn, Coogan:2021uqv, Cole:2022fir, Becker:2022wlo}, which could be detected by future gravitational wave detectors such as the Laser Interferometer Space Antenna (LISA). 
It might also be used to test interactions between dark matter and standard model particles~\cite{Cline:2022qld, Ferrer:2022kei}. 
Interestingly, indirect evidence of the dark matter spike is recently reported from the observations of anomalously fast orbital decays of companion stars near black holes~\cite{Chan:2022gqd}.

Wave dark matter provides an interesting alternative dark matter candidate. 
Wave dark matter refers to a light bosonic dark matter with a mass smaller than a few eV and with a large occupation number.
Wave dark matter arises from various beyond the standard models, for instance, QCD axion models~\cite{Peccei:1977hh, Weinberg:1977ma, Wilczek:1977pj, Preskill:1982cy, Abbott:1982af, Dine:1982ah} for the strong CP problem, dynamical solutions to the electroweak hierarchy problem~\cite{Graham:2015cka, Arvanitaki:2016xds, Banerjee:2018xmn, Banerjee:2020kww, Arkani-Hamed:2020yna, TitoDAgnolo:2021nhd, TitoDAgnolo:2021pjo, Chatrchyan:2022pcb, Chatrchyan:2022dpy}, and ultraviolet theories with compact extradimensions~\cite{Svrcek:2006yi, Arvanitaki:2009fg}.
Due to the large occupation number, wave dark matter candidate behaves closer to classical waves, leading to a number of interesting and qualitatively distinctive phenomenological signatures in the early universe as well as in the late universe. See~\cite{Lee:2017qve, Hui:2021tkt} for recent review.

The aim of this work is two-fold.
First, we investigate how the wave dark matter responds to the adiabatic change of the system, such as the growth of the central black hole or the collapse of molecular clouds at the center of the system.
We show that the adiabatic compression takes place in a similar way as in the particle dark matter halo, while the central density profile in the compressed wave halo is distinctive from that of the particle halo. 
Second, we study the dephasing of gravitational waves from a compact solar-mass object orbiting around the central black hole within the compressed wave halo. 
Gravitational wave emission from the binary within the compressed halo has been studied in the context of compressed particle halo (dark matter spike)~\cite{Eda:2013gg, Eda:2014kra, Kavanagh:2020cfn, Coogan:2021uqv, Cole:2022fir, Becker:2022wlo}. 
We investigate how the microscopical nature of dark matter changes the gravitational wave emission in such compressed halos, paying particular attention to the difference and similarity to the particle halo.
We show that in certain physics scenarios, gravitational wave observations could provide an interesting opportunity to probe wave dark matter around $m \sim 10^{-13}\eV$. 

The paper is organized as follows. 
In section~\ref{sec:ac}, we discuss how the wave dark matter halo responds to the adiabatic change of the system. 
In section~\ref{sec:IMRI}, we study the gravitational wave signals from an inspiral between a central massive black hole and a solar-mass companion object as one example of astrophysical implications of the compressed halo. 
We show that the compressed wave halo, or {\it wave spike}, exerts a dynamical friction force and affects the phase of gravitational waves emitted from a binary embedded in the halo.
We show that such signals are distinctive from other scenarios such as gravitational waves from compressed particle halo and that they can be detectable in future gravitational wave detectors, such as Laser Interferometer Space Antenna (LISA). 
In section~\ref{sec:discussion}, we discuss how the assumptions and approximations that we made in the analysis affect our results.

\section{Adiabatic compression}\label{sec:ac}

\subsection{Review}
Before we study the adiabatic compression of the wave dark matter halo, we review how the particle halo responds to the adiabatic change of the system.
We are interested in scenarios such as the adiabatic growth of a central black hole or the collapse of a molecular cloud through dissipation. 
For the following discussion, we first discuss the phase space distribution of a given density profile, investigate how the phase space distribution evolves under the adiabatic change of the system, and compute the density profile during and after the adiabatic change of the system.

The dark matter density profile is given by the phase space integral of the phase space distribution $f(\bfx,\bfv)$, 
\bea
\rho(\bfx) 
= \int d^3 v \, f(\bfx,\bfv) 
= \int d E \int dL \frac{4\pi L}{r^2 v_r} f(E,L),
\label{classical}
\eea
where $v_r = \sqrt{2(E-\Phi) - L^2/r^2}$ is the radial velocity.
We consider a spherically symmetric halo in this work.
Here $E$ and $L$ denote energy and angular momentum per unit mass. 
In the second expression, we assume that the phase space distribution only depends on the energy $E$ and the absolute value of angular momentum $L = | {\pmb L}|$. 

The above expression can be inverted.
If the phase space distribution depends only on the energy $E$, the phase space distribution for a density profile $\rho(\bfx)$ is~\cite{2008gady.book.....B}
\bea
f(E) = \frac{1}{2\sqrt{2}\pi^2} \frac{d}{dE}
\left( \int^0_E d\Phi\, \frac{d\rho/d\Phi}{\sqrt{\Phi-E} }
\right). 
\label{eddington}
\eea

Suppose now that the system changes slowly:
$$
H_0 = H(t_0) 
\to H_1 = H(t_1). 
$$
By slowly, we mean that the Hamiltonian changes on a time scale longer than the crossing time scale but shorter than the halo relaxation time scale.
Under the adiabatic change of the system, the adiabatic invariants, such as the radial action $J_r = \frac{1}{2\pi} \oint dr \,v_r =  \frac{1}{2\pi} \oint dr\, \sqrt{2(E-\Phi) - L^2/r^2}$ and the angular momentum $L$, are conserved. 
From the conservation of adiabatic invariants, it is straightforward to show that the distribution function at the end of the adiabatic evolution is given by~\cite{1980ApJ...242.1232Y}
\bea
f (E_1, L_1) = f(E_0 (E_1,L_1) )\ .
\label{particle_phase}
\eea
Note that the subscript $0, 1$ denotes quantities at the beginning ($t_0$) and the end ($t_1$) of adiabatic evolution. 
Since the angular momentum is conserved, we find $L_0 = L_1 = L$.
The relation between final energy and initial energy can be found by solving
\bea
J_{r, 1} ( E_1, L_1) = J_{r,0}(E_0, L_0)\ .
\eea
From the above equation, one obtains $E_0 = E_0 (E_1, L_1)$. 
The final mass density is written as
\bea
\rho_1(r)
=
\int_{\Phi_1(r)}^0 
dE_1 \int_{L_{1, \rm min}}^{L_{1,\rm max}}  dL_1
\frac{4\pi L_1}{r^2 v_r} f(E_0 (E_1, L_1) ). 
\label{adiabatic}
\eea
Although the initial distribution does not depend on the angular momentum, the final distribution function depends on the angular momentum in general, causing a mild velocity anisotropy of the compressed dark matter halo~\cite{Merritt:2003qc}.

As an example, let us consider an initial power-law profile $\rho_0(r)= \rho_s ( r_s /r)^\gamma$. 
Suppose that a central Schwarzschild black hole of mass $M_{\rm BH}$ has formed and grown via adiabatic processes. 
The final mass density profile at the end of adiabatic evolution is
\bea
\rho_1 (r) \approx
\rho_{\rm sp}
\left( \frac{ r_{\rm sp} }{ r} \right)^{\gamma_{\rm sp}} 
\left( 1 - \frac{2R_S}{r} \right)^{\gamma_{\rm sp}},
\label{particle_halo}
\eea
where $\gamma_{\rm sp} = (9 -2\gamma) / (4-\gamma)$~\cite{1995ApJ...440..554Q, Gondolo:1999ef}, $\rho_{\rm sp} = \rho_0(r_{\rm sp})$, and $R_S =  2 GM_{\rm BH}$. 
Since $\gamma_{\rm sp} > \gamma$, the adiabatic compression provides a steeper dark matter profile.
The above profile is valid up to $r_{\rm sp} = 0.2 r_h$
\bea
r_{\rm sp} 
=  \left[ \frac{(3-\gamma_{\rm sp}) 0.2^{3-\gamma_{\rm sp}} M_{\rm BH} }{2\pi \rho_{\rm sp}} \right]^{1/3}
\label{rsp}
\eea
where $r_h$ is defined as a radius such that the enclosed mass is twice the central black hole mass, $M_{\rm enc}(r<r_h) = 2 M_{\rm BH}$~\cite{Merritt:2003qc}.  
The profile converges to the initial profile, $\rho_1 (r) \simeq \rho_0(r)$, for $r > r_{\rm sp}$. 
An additional factor $(1-2R_S/r)^{\gamma_{\rm sp}}$ is introduced as the central black hole inevitably absorbs particles with angular momentum smaller than $L \simeq 2 R_S$. 
This approximate profile matches to relativistic results in Sadeghian et al~\cite{Sadeghian:2013laa} with an error less than $20\%$ over all radii.

\subsection{Construction of wave DM halo}\label{sec:wave_DM_halo}
We now investigate how the wave dark matter halo responds to the adiabatic change of the system.
The key observation from the previous section is the conservation of adiabatic invariants and the relation between initial and final phase space distribution Eq.~\eqref{particle_phase}. 
To see how it can be generalized to the discussion of wave halo, we first discuss the initial wave dark matter halo profile and its decomposition into eigenmodes, which provides a description corresponding to the phase space distribution in the particle limit. 
Then, we use the Schr{\"o}dinger equation and the adiabatic theorem to prove that the response of wave dark matter is similar to that of particle dark matter in the semiclassical limit. 
Near the center of the compressed wave halo, where we cannot use the semi-classical approximation, there could be a characteristic core, whose profile is given by the ground state solution of a hydrogen-like atom. 
This core may or may not survive over the astrophysical time scale depending on the dark matter and the central black hole mass, which will be discussed shortly. 

\subsubsection{Density profile}
Numerical simulations of ultralight dark matter found that the wave dark matter halo consists of an NFW-like profile and soliton core.
We assume that the wave halo is given by
\bea
\rho (r) = 
\begin{cases}
\rho_{\rm c}(r) & r < r_t
\\
\rho_{0}(r) & r \geq r_t
\end{cases}
\label{initial_profile}
\eea
where $\rho_c(r)$ is the core profile, $\rho_{0}(r)$ is some unspecified NFW-like profile and $\rho_0 (r_t) = \rho_c(r_t)$.
The core density $\rho_c(r)$ and the core mass $M_c$ are~\cite{Schive:2014hza}
\bea
\rho_{\rm c}(r) &\simeq& \frac{2 M_\odot /{\rm pc}^3}{[ 1 + 0.091 (r/r_c)^2]^8}  \bigg( \frac{\rm kpc}{r_c} \bigg)^4
\bigg( \frac{10^{-23}\eV}{m} \bigg)^2 ,
\label{rho_c}
\\
M_c &=& 6\times 10^9 M_\odot 
\bigg( \frac{10^{-23}\eV}{m} \bigg)^2 \bigg( \frac{\rm kpc}{r_c} \bigg) .
\eea
The core mass $M_c$ is defined as the mass enclosed within the core radius $r_c$, which is defined as $\rho_c(0) =2 \rho_c(r_c)$.
This relation holds as long as the central core object is the ground state solution to the Schr{\"o}dinger-Poisson equation.
The total soliton mass is related to the core mass as $M_c \simeq 0.24M_{\rm sol}$. 

The core radius $r_c$ is a free parameter. It is due to the scaling symmetry of the Schr{\"o}dinger-Poisson equation. 
We determine the core radius as
\bea
r_c = \frac{c_r}{mv_c(r_c)}
\label{core_radius}
\eea
with some constant $c_r\sim{\cal O}(1)$. 
Here $v_c$ is the circular velocity. 
This implicit equation fixes the core radius to the typical wavelength of dark matter. 

The core radius connects the property of the host halo to the soliton radius. 
The easiest way to see this is by rewriting the expression as $r_c = c_r^2 / GM_{\rm enc}(r_c) m^2$.
The ground state solution to the Schr{\"o}dinger-Poisson equation predicts $r_c = 2.67 / GM_{\rm sol} m^2$~\cite{Chavanis:2011zi, Hui:2016ltb}. 
Imposing the core radius relation~\eqref{core_radius} is equivalent to say that the soliton mass is determined by the enclosed mass of halo within the typical wavelength of dark matter. 

Numerical simulations have also found a so-called the soliton-host halo relation~\cite{Schive:2014dra, Schive:2014hza}, relating the properties of host halo to the soliton properties. 
This soliton-host halo relation can be written as $G M_c m = (E/M)^{1/2}_{\rm halo}$~\cite{Schive:2014dra, Schive:2014hza} or equivalently as $(K/M)_{\rm sol}^{1/2} = (K/M)^{1/2}_{\rm halo}$~\cite{Bar:2018acw, Bar:2019bqz, Bar:2021kti}, where $(E/M)_{\rm halo}$ is the energy of halo per unit mass, and $(K/M)_{\rm sol}$ and $(K/M)_{\rm halo}$ are the kinetic energy per unit mass in soliton and in halo, respectively.
These can be written as 
\bea
r_c \sim \frac{1}{mv_{\rm vir}}. 
\label{soliton_hosthalo}
\eea
Comparing to Eq.~\eqref{core_radius}, the only difference is that the circular velocity is being replaced with the virial velocity of the halo.\footnote{Note that the soliton-host relation has been debated. See~\cite{Chan:2021bja} for a recent review. 
See also~\cite{Bar:2019pnz} for the discussion on the validity of soliton-host halo relation for larger particle mass.}

The difference between \eqref{core_radius} and \eqref{soliton_hosthalo} is negligible as long as the size of the system is comparable to the wavelength of dark matter.
Stating it differently, the circular velocity at the core radius is similar to the virial velocity if the core radius is not much smaller than the scale radius of the halo. 
This is indeed the case for most of numerical simulations, e.g.~\cite{Schive:2014dra, Schive:2014hza, Mocz:2017wlg, Veltmaat:2018dfz}. 

A subtlety arises when the wavelength of dark matter is hierarchically smaller than the size of the system.
In this case, the wave dark matter can probe the inner part of the halo, and a direct application of usual soliton-host halo relation~\eqref{soliton_hosthalo} could underestimate the size of the soliton; it may significantly overestimate the soliton mass~\cite{Bar:2019pnz}. 
On the other hand, the relation~\eqref{core_radius} predicts a smooth transition from NFW-like profile to cored profile when the radius is around the wavelength of dark matter. 
Since we are interested in a wide range of wave dark matter mass, for instance up to $m\simeq 10^{-13}\eV$, we use the core radius~\eqref{core_radius} for our analysis.

\subsubsection{Eigenmode decomposition}
Having discussed the density profile of wave dark matter, we consider an effective description of wave dark matter halo with eigenmode decomposition, following previous works by Lin et al~\cite{Lin:2018whl} and Yavets et al~\cite{Yavetz:2021pbc}.
The occupation number for each eigenmode plays a similar role as the phase space distribution in the particle halo. 

We begin with an action of a massive scalar field 
\bea
S = \int d^4x \, \sqrt{-g} 
\left[ \frac{1}{2} g^{\mu\nu} \partial_\mu \phi \partial_\nu \phi 
 - \frac{1}{2} m^2 \phi^2\right]
\eea
where $g_{\mu\nu}$ is the metric tensor. 
The line element is $ds^2 = (1+2 \Phi)dt^2 - (1-2\Phi) \delta_{ij} dx^i dx^j$ with the gravitational potential $\Phi$. 
We ignore the self-interaction in this work. 
In the non-relativistic limit, we expand the field as
$$
\phi(x,t) = \sum_i \frac{1}{\sqrt{2m}} 
\left[ a_i  \psi_i(x,t) e^{-imt} + a_i^\dagger \psi_i^*(x,t) e^{+ imt} \right],
$$
where $a_i$ and $a_i^\dagger$ are annihilation and creation operator, and $\psi(x,t)$ is the wavefunction that evolves according to the Schr{\"o}dinger-Poisson equation, 
\begin{equation}
\label{eq:SP}
\begin{aligned}
i \hbar \frac{\partial \psi}{\partial t} &=\left(-\frac{\hbar^2 \nabla^2}{2 m} + m \Phi \right) \psi ,
\\
\nabla^2 \Phi
&= 4 \pi G m |\psi|^2.
\end{aligned}
\end{equation}
The wavefunction is normalized as $1 = \int d^3x \, |\psi(x,t)|^2$. 
In the wave limit, the operator $a_i$ can be thought as a commuting random variable $\alpha_i$ whose probability density function 
$p(\alpha_i)$ is given by~\cite{Kim:2021yyo},
\bea
p(\alpha_i) = \frac{1}{\pi f_i}
\exp\left[ - \frac{ |\alpha_i|^2 }{ f_i } \right]
\label{pdf_alpha}
\eea
where $f_i$ is the mean occupation number of the mode $i$. 
The modulus $|\alpha_i|$ follows the Rayleigh distribution; the phase ${\rm arg}(\alpha_i)$ follows the uniform distribution.

We model the wave dark matter halo as a superposition of eigenmodes of the Schr{\"o}dinger-Poisson equation~\eqref{eq:SP}, 
\bea
\langle \rho \rangle \equiv \bar \rho = m \sum_i f_i | \psi_i(x) |^2 ,
\eea
where $f_i$ is the occupation number for each mode $\psi_i$ and the angle bracket denotes an ensemble average with respect to the probability density $p(\alpha_i)$. 
To find the eigenmodes, we assume time-independent gravitational potential $\Phi_t$ obtained from the Poisson equation $\nabla^2\Phi_t = 4\pi G\rho_t$, where $\rho_t$ is an initial target density profile given by the soliton and NFW-like profile~\eqref{initial_profile}.  
The eigenmodes then can be found by solving the time-independent Sch{\"o}dinger equation,
\bea
\label{eq:EigenMode Schrodinger}
0  = \Big[ \nabla^{2} -  2m^2 (\Phi_t - E_i)  \Big] \psi_{i},
\eea
where $E_i$ is the energy eigenvalue divided by the particle mass. 
We find the occupation number for each mode $f_i$ such that $\bar\rho \approx \rho_t$.
Practically, this can be done by minimizing the cost function
$D(\rho_t, \bar \rho) = \frac{1}{r} \int^r_0 dr \, (\rho_t - \bar \rho)^2 / \rho_t^2$ as discussed in Yavetz et al~\cite{Yavetz:2021pbc}. 
To find a self-consistent profile, we iterate the above fitting procedure several times by updating the target density profile and gravitational potential until the density profile converges.

This construction is qualitatively identical to the particle halo construction with phase space distribution. 
In the particle halo, we assume that the DM particle is distributed according to the phase space distribution, $f(\bfx,\bfv)$, where each trajectory evolves according to the Hamiltonian of the system, which is constructed with the mean potential of the system $\Phi$. 
The density profile $\rho = \int d^3v \,f(\bfv,\bfx)$ should then satisfy the Poisson equation, $\nabla^2 \Phi = 4\pi G \rho$, for the self-consistency. 
In the wave dark matter halo, we follow the same procedure with the additional step of solving the Schr{\"o}dinger equation to find eigenfunctions.

Since we consider a spherically symmetric system, it is more convenient to decompose the wavefunction into a radial wavefunction and spherical harmonics,
\bea
\psi_i(\bfx) \to \psi_{n\ell m_\ell}(\bfx) =R_{n\ell}(r) Y_\ell^{m_\ell}(\theta,\phi)
= \frac{\chi_{n\ell}(r)}{r} Y_{\ell}^{m_\ell} (\theta,\phi)
\nonumber
\eea
where the radial wave function satisfies $\chi''_{n\ell} + [2m^2(E_{n\ell} - \Phi) - \ell (\ell + 1 )/r^2 ] \chi_{n\ell} = 0$. 
The mean density is then given by
\bea
\bar\rho (r)
= \frac{m}{4 \pi} \sum_{n \ell} (2 \ell+1) f_{n \ell} |R_{n\ell}(r)|^2
\label{wave_rho}
\eea
where we assume that the occupation number does not depend on the magnetic quantum number as we consider a spherically symmetric system. 

For the following discussion, it is useful to see how the above wave dark matter construction~\eqref{wave_rho} compares to the particle dark matter~\eqref{classical}. 
The correspondence between particle and wave halo becomes clearer in the semi-classical limit~\cite{Yavetz:2021pbc}. 
We may approximate the discrete sum in Eq.~\eqref{wave_rho} with the continuous integral as
\bea
\bar\rho 
\approx
\int dE \int dL \, \frac{4\pi L }{v_r r^2} 
\left[ \frac{m^4}{(2\pi)^3}  f_{n\ell} \right] 
\left( \frac{\pi}{m} \frac{dn}{dE} \right) (v_r |\chi_{n\ell}|^2),
\nonumber
\eea
where $dn/dE$ is the density of the state. 
In the semiclassical limit $mvr \gg 1$, the WKB approximation can be used to find a radial wavefunction as
\bea
\chi_{n\ell} 
=  \frac{ {\cal N}_{n\ell}}{ \sqrt{m v_r(r)} }
\cos \left[  m \int^r_{r_{\rm min}} dr' \, v_r(r') - \frac{\pi}{4} \right]
\label{a}
\eea
with the normalization constant
\bea
|{\cal N}_{n\ell}|^2 
\approx
\left[ \int_{r_{\rm min}}^{r_{\rm max}} \frac{dr}{2mv_r(r)} \right]^{-1} .
\eea
Due to the boundary condition, one obtains the Bohr-Sommerfeld semi-classical quantization condition,
\bea
(n+ 1/2) \pi = m \int_{r_{\rm min}}^{r_{\rm max}} dr' \, v_r(r'),
\label{BS_quantization}
\eea
from which one can compute the density of the states, 
\bea
\frac{dn}{dE} = \frac{m}{\pi} \int_{r_{\rm min}}^{r_{\rm max}} \frac{dr}{v_r(r)} .
\eea
Here $r_{\rm max}$ and $r_{\rm min}$ are turning points at which $v_{r}(r) = 0$. 
Combining these expressions, we find
\bea
\bar\rho
\approx
\int dE \int dL \, \frac{4\pi L }{v_r r^2} 
\left[ \frac{m^4}{(2\pi)^3}  f_{n\ell} \right] 
\eea
where we have replaced $2 \cos^2 [ m \int^r dr' \, v_r(r')  - \pi/4]  \approx 1$. 
The correspondence between the occupation number and classical phase space distribution is found as~\cite{Yavetz:2021pbc}
\bea
f(E) \simeq \frac{m^4}{(2\pi)^3} f_{n\ell}. 
\eea
This allows us to interpret the occupation number as a phase space distribution in the semiclassical limit.

\begin{figure}
\centering
\includegraphics[width=0.45\textwidth]{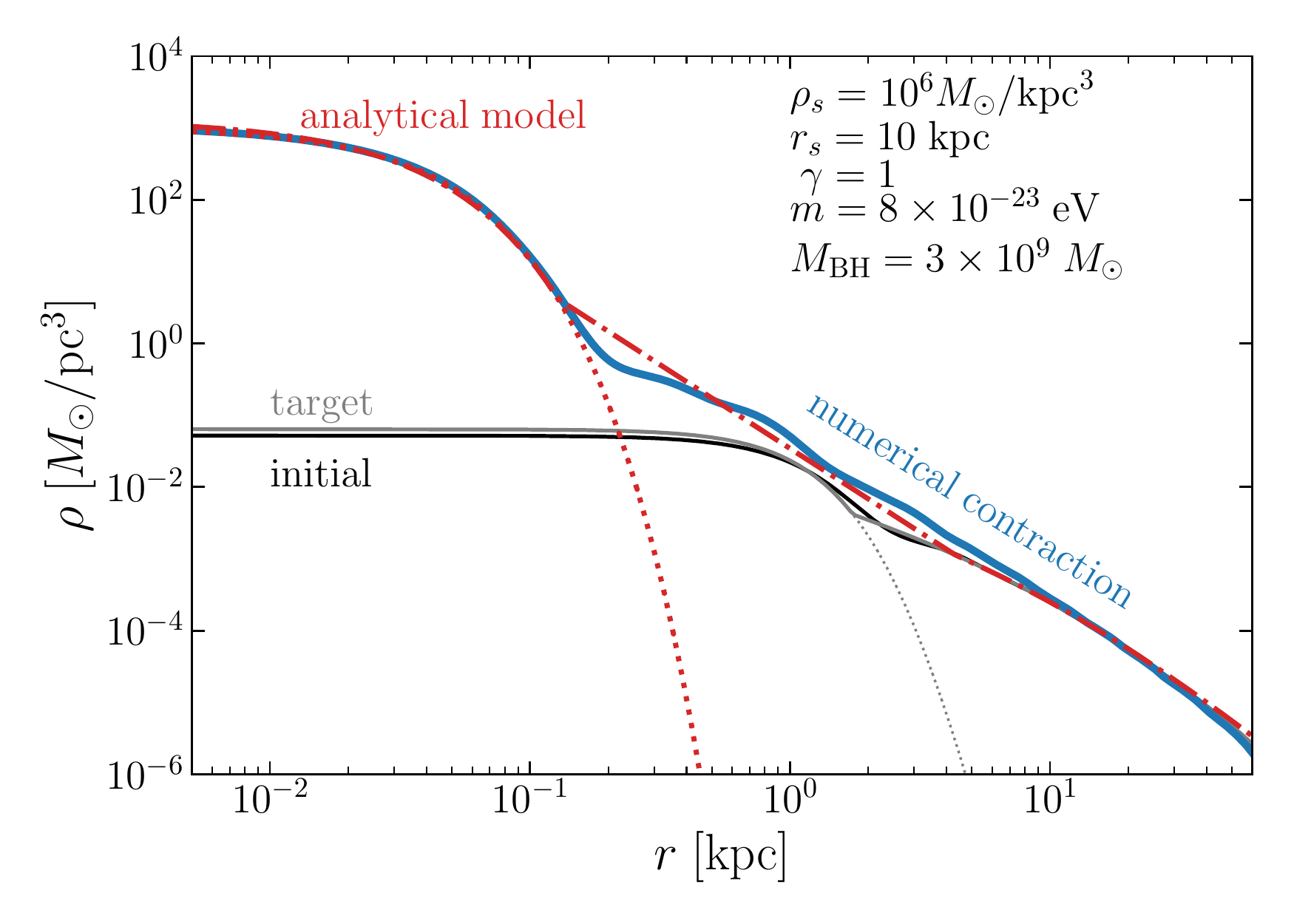}
\\
\includegraphics[width=0.45\textwidth]{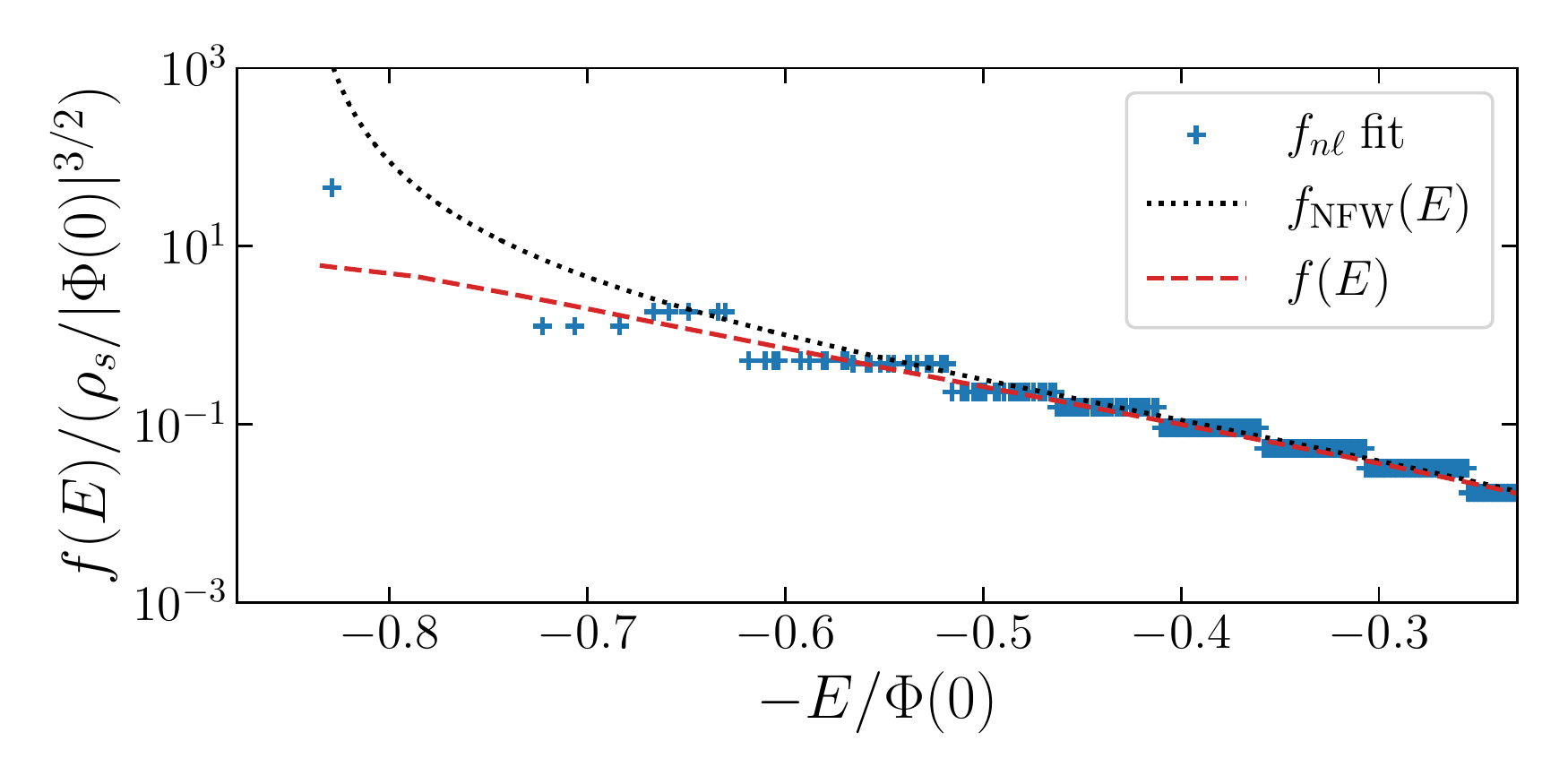}
\caption{(Top) the initial and compressed wave halo profile for $m = 8 \times 10^{-23}\eV$.
The initial target profile is given by Eq.~\eqref{initial_profile}, where $\rho_0 (r) = \rho_s / [(r/r_s)(1+r/r_s)^2]$ with $\rho_s = 10^{-3} M_\odot/{\rm pc}^3$ and $r_s = 10\,{\rm kpc}$ (gray). 
The black solid line is the density profile obtained by fitting the occupation numbers such that $\bar\rho \simeq \rho_t$.
The blue line is the compressed wave halo, which consists of the central solitonic core and outer spike profile that is similar to the particle dark matter halo. 
For this, we assume an adiabatic formation of the central black hole of mass $M_{\rm BH} = 3\times 10^9 M_\odot$.
(Bottom) 
The dotted line is the phase space distribution of NFW profile, the dashed line is the phase space distribution of target density profile obtained by the inversion formula~\eqref{eddington}, and the crosses are the corresponding occupation numbers for wave halo $[m^4/(2\pi)^3] f_{n\ell}$.
}
\label{fig:compression_wave}
\end{figure}

In Figure~\ref{fig:compression_wave}, we show the density profile constructed according to the method described above as well as the occupation number. 
As an example, we consider an initial soliton and NFW profile (gray line) with scale density $\rho_s = 10^6 \, M_\odot/{\rm pc}^3$, the scale radius $r_s= 10 \, {\rm kpc}$, and particle mass $m = 8\times 10^{-23}\eV$.
We then obtain the gravitational potential from this target density profile, solve the Schr{\"o}dinger equation, and find a set of occupation numbers for each mode that reproduces the initial profile closest. 
More specifically, we bin energy eigenvalues and assign the same occupation number to modes that have the energy eigenvalue in the same energy bin.
We iterate the procedure three times to ensure the self-consistency of the halo.
The resulting halo constructed from the superposition of eigenmodes is shown as a black line. 
The corresponding occupation number is shown in the bottom panel of Figure~\ref{fig:compression_wave}.
As expected, the occupation number $f_{n\ell} [m^4/(2\pi)^3]$ closely follows the classical phase space distribution $f(E)$ except for the ground state. 
The ground state occupation number is simply given by $f_0 \simeq M_{\rm sol} / m$ with the total soliton mass $M_{\rm sol}$.

\subsection{Compression of wave DM halo}\label{sec:wave_compression}
We now consider the adiabatic change of the system.
Suppose that the initial Hamiltonian $H_0 = H(t_0)$ adiabatically changes to $H_1 = H(t_1)$ as before. 
According to the adiabatic theorem, an eigenstate of the Hamiltonian at an initial time ($t_0$) remains as an instantaneous eigenstate of the Hamiltonian as the system evolves. 
If the initial density profile is given by $\bar\rho_0(r) = m \sum_i f_i |\psi^{(0)}_i (r)|^2$ where $\psi^{(0)}_i$ is the eigenmode of the initial Hamiltonian $H_0$, the final density profile is given by
\bea
\bar\rho_1(r)
&=& m \sum_i f_i |\psi^{(1)}_i (r)|^2
\nonumber\\
&=& \frac{m}{4 \pi} \sum_{n\ell} (2\ell+1) f_{n\ell} |R_{n\ell}^{(1)}(r)|^2
\label{final_profile}
\eea
with the eigenmode $\psi^{(1)}_i = R_{n\ell}^{(1)} Y_\ell^{m_\ell}$ of the final Hamiltonian $H_1$. 
The occupation number remains the same, $f_{n'\ell'} = f_{n\ell}$, throughout the adiabatic evolution.

To see how this compressed wave halo compares to the compressed particle halo, we consider the semiclassical limit, $mv r \gg 1$. 
In the semiclassical limit, the wave dark matter density profile is approximated as $\rho = \int dE dL \frac{4\pi L}{v_r r^2} f(E,L)$ with a substitution $f(E,L) \to [m^4/(2\pi)^3]f_{n\ell}$ and $L^2 = \ell(\ell+1)/m^2$. 
Since the eigenstate stays in the same eigenstate during the adiabatic evolution, the quantum number before and after the adiabatic evolution is the same; $n' = n$ and $\ell' = \ell$. 
The conservation of quantum numbers is translated into the conservation of classical adiabatic invariants.
For instance, the condition $\ell' = \ell$ leads to $L_0 = L_1 = L$, while the condition $n' = n$ leads to 
\bea
\oint dr' v_{r, 0}(r')  = \oint dr' v_{r, 1}(r') 
\eea
from the semi-classical Bohr-Sommerfeld quantization condition~\eqref{BS_quantization}. 
Since the radial action is defined as $J_r = \frac{1}{2\pi} \oint dr\, v_r(r)$, this condition is identical to the conservation of the radial action, $J_{r, 1}(E_0,L_0) = J_{r, 1}(E_1, L_1)$, which we use to derive $E_0 = E_0(E_1 , L_1)$ for the particle halo compression. 
From these arguments, we see that the adiabatic compression proceeds in the same way as in the particle case as long as the semiclassical approximation is valid. 

The semiclassical approximation breaks down at small $r$.
At small radii, the profile is dominated by low angular momentum modes, and especially the ground state solution.
Assuming that the ground state dominates the central profile, the core profile before the adiabatic compression is given by
\bea
\rho_{c,0} 
= m f_{0} |\psi^{(0)}_{0}(r)|^2 
= M_{\rm sol}|\psi^{(0)}_{0}(r)|^2.
\eea
where $\psi^{(0)}_{0}$ is the ground state wave function of initial system $H_0$ and $f_0 = M_{\rm sol}/m$ is the occupation number for the ground state. 
The above form is a simple reparametrization of the soliton profile $\rho_c(r)$~\eqref{rho_c}. 
Since the occupation number is conserved, the final core profile is
\bea
\rho_{c,1}
 = m f_0 |\psi^{(1)}_{0}(r)|^2
 = M_{\rm sol} |\psi^{(1)}_{0}(r)|^2.
\eea
Here $\psi^{(1)}_{0}$ is the ground state of the final Hamiltonian $H_1$. 
If the central black hole dominates the gravitational potential of the final system, 
the Hamiltonian $H_1$ can be approximated as 
$$H_{1} \approx \frac{p^2}{2m} - \frac{ GM_{\rm BH} }{r},$$
and, therefore, the ground state wave function is given by the ground state wave function of the hydrogen atom,
\bea
\psi_0(r) = \frac{e^{-r/a}}{\sqrt{\pi} a^{3/2}} 
\eea
with a gravitational Bohr radius $a = 1/ (m \alpha_G ) =1 / ( G M_{\rm BH} m^2)$.
The final compressed wave dark matter halo consists of the central soliton and the outer cuspy profile 
\bea
\rho_1 (r)
=
\begin{cases}
\rho_{c,1}(r) & r < r_t,
\\
\rho_{\rm sp}(r_{\rm sp}/r)^{\gamma_{\rm sp}} &  r_t < r <r_{\rm sp} , 
\end{cases}
\label{compress_analytic}
\eea
where $r_t$ is defined such that $\rho_{\rm sp}(r_{\rm sp}/ r_t)^{\gamma_{\rm sp}} = \rho_{c,1}(r_t)$. 
The profile for $r > r_t$ is the same as the compressed particle halo, $ \rho_{\rm sp} ( r_{\rm sp} /r )^{\gamma_{\rm sp}}$, which is valid up to $r \simeq r_{\rm sp}$. 
In summary, the central part is replaced by the soliton and the outer part is the same as compressed particle dark matter halo. 

In the top panel of Figure~\ref{fig:compression_wave}, we show the compressed wave halo by solving the Schrodinger equation of the final Hamiltonian $H_1$ with the occupation number obtained in the previous section.
For the illustration, we choose the central black hole mass $M_{\rm BH} =3 \times 10^9M_\odot$.
The resulting compressed profile (blue solid) consists of the central core, and outer spike profile similar to the compressed particle halo. 
The central density profile is given by the hydrogen ground state wave function. 
We also show the analytical profile given in Eq.~\eqref{compress_analytic} (red dashed), which agrees with the numerical result. 
To ensure the self-consistency of the compressed halo, we also perform three iterations to obtain this result. 

\subsection{Survival of the core}
If a massive black hole is at the center of halo, it may swallow the low angular momentum modes over the astrophysical time scale. 
It is exactly this reason why the particle halo density profile sharply drops at $r = 2R_s$. 

The central black hole can absorb wave modes in a similar fashion. 
Whether a certain wave mode can survive depends on the central black hole mass and the wave dark matter mass. 
To investigate this, we take the Schwarzschild black hole for simplicity.
While the system is identical to the hydrogen atom in the non-relativistic limit, the energy eigenvalue develops an imaginary part due to the boundary condition at the black hole horizon. 
The imaginary part of the energy eigenvalue is interpreted as a decay rate and is given by
\bea
\frac{ \Gamma_{n\ell} }{ m } \approx
\alpha_G^{4\ell +5} 
\frac{2^{4\ell+3}(n+\ell)!}{n^{2\ell+4} (n-\ell-1)!} 
\left[
\frac{\ell !}{(2 \ell)! (2\ell+1)!} 
\right]^2
\nonumber\\
\times
\prod_{k=1}^\ell [k^2 + (4\alpha_G)^2]
\label{decay_rate}
\eea
where we take the Schwarzschild limit of the result obtained in the Kerr geometry~\cite{Detweiler:1980uk, Baumann:2019eav}. 
Here $\alpha_G = G M_{\rm BH} m$. 
Including the possible decay of modes into the black hole, the wave dark matter halo density at a given time is
\bea
\bar \rho(r) = \frac{m}{4\pi} \sum_{n\ell} (2\ell+1) |R_{n\ell}(r)|^2 e^{-2 \Gamma_{n\ell} t_{\rm halo}} .
\label{halo_w_decay}
\eea
where $t_{\rm halo}$ is the age of the dark matter halo. 
Compared to the previous wave dark matter profile, we have an additional exponential decay factor due to the absorption. 
The above estimation assumes that the central black hole dominates the dynamics of the system.
We only consider such cases in this work.
The decay of the central soliton given by Eq.~\eqref{decay_rate} agrees with full numerical simulations~\cite{Cardoso:2022nzc}.

The decay rate of the ground state scales as $\Gamma \propto m \alpha_G^5$. 
For the example shown in Figure~\ref{fig:compression_wave}, the gravitational fine structure constant is $\alpha_G \sim 2\times 10^{-3}$ and the decay rate is $\Gamma \sim (10^3\,{\rm Gyr})^{-1}$; the solitonic core of the compressed wave halo survives over the age of the universe in this case. 
Note, however, that whether the soliton and low angular momentum modes can survive over the age of the halo crucially depends on $\alpha_G = G M_{\rm BH} m $. 
In the example to be discussed in the following section, low angular momentum modes are indeed unstable over the age of the halo as the central black hole absorbs them.
The decay of wave modes leads to a distinctive compressed wave halo  compared to the one studied above. In particular, it can be described with a broken power-law profile, which will be discussed in the next section.

\section{Application: Intermediate-Mass Ratio Inspirals}\label{sec:IMRI}
The adiabatic growth of the black hole concentrates the dark matter mass density near the center. 
As discussed in previous works (e.g. Ref.~\cite{Eda:2013gg}), a compressed dark matter profile could lead to the dephasing of gravitational wave signals from a binary system, which could be observed by future gravitational wave detectors such as LISA. 
In this section, we investigate an intermediate-mass ratio inspiral -- an inspiral between an intermediate-mass black hole ($m_1$) at the center of the halo and a solar-mass compact object ($m_2$) -- to see if one can extract information on the surrounding wave dark matter halo from gravitational wave observations. 

We consider a halo formed at $z=20$ with the virial mass $M_{\rm vir} \simeq 10^6 M_\odot$, which was considered in Eda et al~\cite{Eda:2014kra}.
We model the initial (outer) profile as
\bea
\rho_0(r)  = \frac{\rho_s}{(r/r_s)^\gamma(1 + r/ r_s)^{3-\gamma} } 
\label{benchmark_halo}
\eea
with the scale density $\rho_s \simeq 5.3 M_\odot/{\rm pc}^3$ and the scale radius $r_s \simeq 23\,{\rm pc}$. 
We consider an intermediate-mass black hole of mass $m_1 = 10^{3} \, \textrm{--} \, 10^5M_\odot$ formed at the center of a halo via the collapse of Population III stars~\cite{Madau:2001sc} or a direct collapse of a gas cloud~\cite{Koushiappas:2003zn}, where their formation compresses the surrounding dark matter halo as described in the previous section.\footnote{See the review~\cite{Greene:2019vlv} for details on the formation of intermediate-mass black holes.
Note that the compressed halo might be compromised due to off-center formation, merger events, and DM-star interactions~\cite{Ullio:2001fb, Merritt:2002vj, Merritt:2003qk}. We refer readers to Ref.~\cite{Bertone:2005xz, Bertone:2009kj} for detailed  discussion on this and for an estimate on the number of compressed halos without major merger events.}
The radius of gravitational influence of such intermediate black holes is smaller than $r_s$, and therefore, the adiabatic compression takes place in the inner part of the halo where the density profile is well approximated by a single power-law, $\rho_0(r) \approx r_s (r_s/r)^\gamma$. 
Such intermediate-mass black holes could be seeds for supermassive black holes observed at high redshifts~\cite{Fan:2006dp, Mortlock:2011va, Banados:2017unc}. 

After the adiabatic compression, a dark matter spike has formed. 
For the particle halo and also for the outer part of the wave halo, the density profile is given as $\rho_1(r) \approx \rho_{\rm sp} ( r_{\rm sp} / r)^{\gamma_{\rm sp}}$ with $\gamma_{\rm sp} = (9-2\gamma) / (4-\gamma)$. 
The spike density and radius are 
\bea
\rho_{\rm sp} &=& \rho_s \left[ \frac{2\pi \rho_s r_s^3}{(3-\gamma_{\rm sp}) 0.2^{3-\gamma_{\rm sp}} m_1} \right]^{\frac{\gamma}{3-\gamma}},
\\
r_{\rm sp} &=& r_s \left[ \frac{(3-\gamma_{\rm sp}) 0.2^{3-\gamma_{\rm sp}} m_1}{2\pi \rho_s r_s^3} \right]^{\frac{1}{3-\gamma}},
\eea
which can be easily derived from $\rho_{\rm sp} = \rho_0(r_{\rm sp})$ and $r_{\rm sp} = [ (3-\gamma_{\rm sp}) 0.2^{3-\gamma_{\rm sp}} m_1 / (2\pi \rho_{\rm sp}) ]^{1/3}$.

\begin{figure}
\centering
\includegraphics[width=0.45\textwidth]{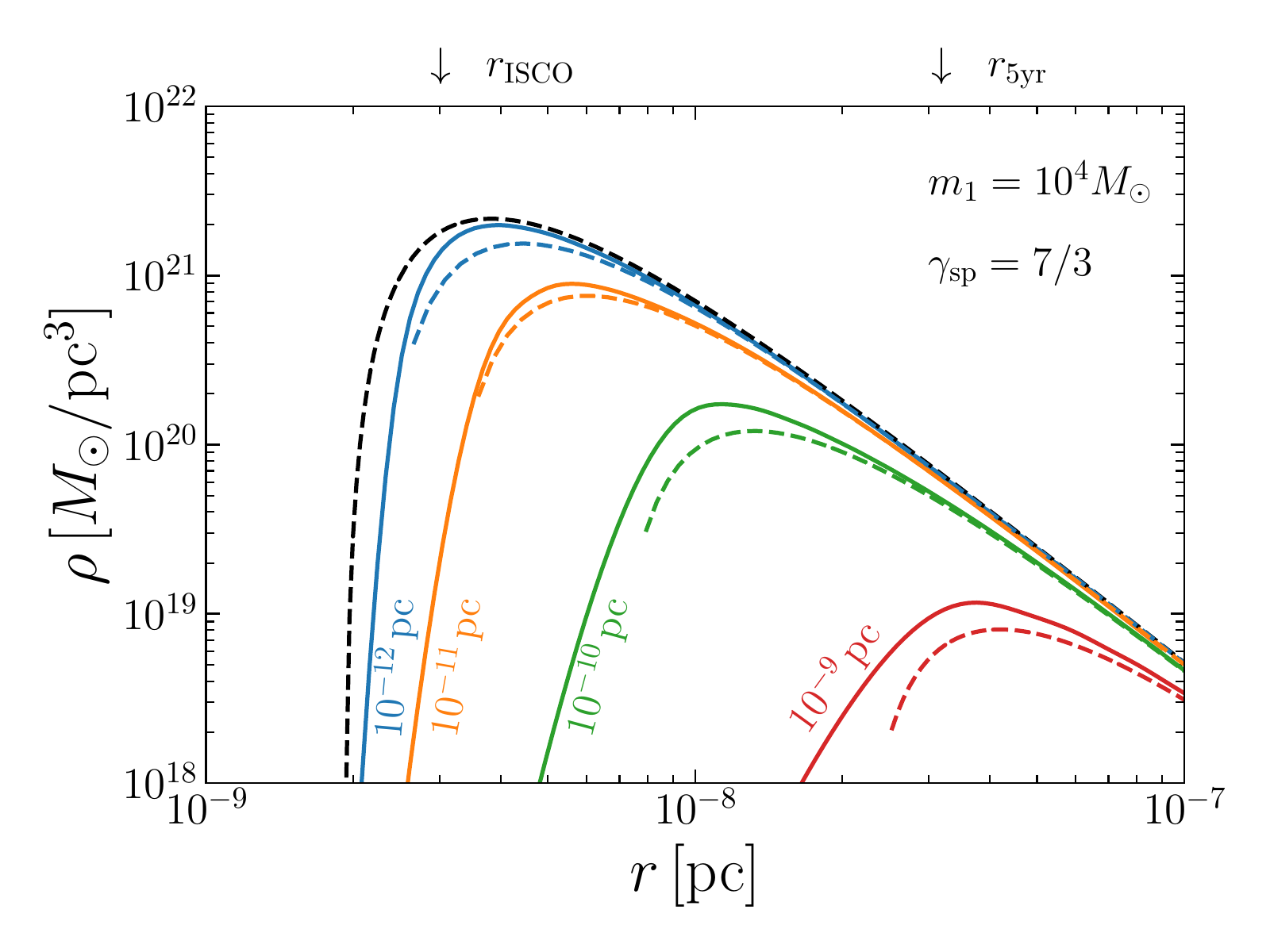}
\caption{The density profile of compressed wave halo. 
The soliton as well as other low angular momentum modes are absorbed by the central black hole, leading to a broken power-law profile. 
At large radii, the profile behaves as $\rho \propto r^{-\gamma_{\rm sp}}$ while at small radii, it behaves as $\rho \propto r^{2\ell_c}$. 
The black dashed line is the profile of particle dark matter halo Eq.~\eqref{particle_halo}. Colored lines show the compressed wave halo profiles for different values of the Bohr radius $a = 1/ G m_1 m^2$.
The dashed lines are the analytical approximation given in Eq.~\eqref{wave_approx}.
Here $r_{5\rm yr}$ is the radial position of a solar-mass compact object at which it coalesces with the central black hole in five years.}
\label{fig:wave_profile}
\end{figure}

Consider now the inspiral between the central intermediate-mass black hole and a solar-mass object in a quasi circular orbit within the compressed wave halo.\footnote{See Ref~\cite{Becker:2021ivq} for discussion on circularization of the orbit with dynamical friction.}
The change of the orbital energy of the companion object is
\bea
\label{eq:energy_loss}
\frac{d E_o}{dt} = 
- \Big( \frac{dE_{\rm GW}}{dt} + \frac{d E_{f}}{dt} \Big),
\eea
where $E_o = -G m_1 m_2 /2r$ is the orbital energy, $m_1$ is the mass of the central black hole, and $m_2$ is the companion mass.

The first term $dE_{\rm GW} / dt$ represents the energy loss due to gravitational wave emission. 
It is given as
\bea
\frac{dE_{\rm GW}}{dt} =  \frac{32}{5G} (G M_c \pi f_{\rm GW} )^{10/3} ,
\eea
where the chirp mass is $M_c = \mu^{3/5} M^{2/5}$ with the reduced mass $\mu = m_1 m_2 / (m_1 + m_2)$ and the total mass $M = m_1 + m_2$.
The gravitational wave frequency is
\bea
f_{\rm GW}  = 2 f_o = \frac{1}{\pi} \sqrt{\frac{GM}{r^3}} 
\eea
with the orbital frequency $f_o = v_c / 2\pi r$. 

The second term $dE_f/dt$ is due to the friction induced by the dark matter halo, called dynamical friction. 
The dynamical friction energy loss can be written as
\bea
\frac{d E_f}{dt} \approx \frac{4\pi (G m_2)^2 \rho(<v)}{v} C(v)
\label{Efdot}
\eea
where $v = \sqrt{GM /r}$ is the circular velocity of the companion object and $\rho(<v) = 4\pi \int_0^v dv' \, v'^2 f(v') \simeq 0.6 \times \rho(r)$ is the mass density of particle whose velocity is smaller than the circular velocity. 
The microscopic nature of dark matter affects both the density $\rho(r)$ and the Coulomb logarithm factor $C(v)$.
It is this dynamical friction energy loss that can potentially allow us to probe the nature of dark matter. 
We are, therefore, interested in how the dynamical friction changes the waveform of gravitational waves during five years of inspiral before the coalescence.  

\subsection{Compressed wave halo}
To compute the dynamical friction force, let us revisit the compressed halo density profile. 
In the previous section, the solitonic core survives for an astrophysical time scale, and the final density profile consists of the ground state of a hydrogen-like atom (soliton) and the outer dark matter spike.
For the benchmark halo we study in this section, the conclusion is different: the central soliton as well as low angular momentum modes cannot survive over an astrophysical time scale, and the profile is approximately described by a broken power-law instead of a spike profile with a characteristic solitonic core. 

To illustrate this, let us consider the wave dark matter mass around $m \simeq 10^{-14}\eV$. 
The Bohr radius is $a = 1 / G m_1 m^2 \sim 10^{-8}\,{\rm pc}$, which roughly coincides with the radial position of the companion ($m_2$) five years before the coalescence. 
In this case, the gravitational fine structure constant is $\alpha_G = 0.3 \times (m_1  / 10^3 M_\odot) (m / 10^{-14}\eV)$, while the decay rate of the ground state is $\Gamma \sim m \alpha_G^5$. 
It is straightforward to see that the ground state decays and cannot survive over the Gyr time scale; the central black hole swallow the whole soliton in a short time scale. 
This conclusion holds for the parameter ranges of interest: the mass of the central black hole, $m_1 \sim 10^{3} \, \textrm{--} \, 10^5M_\odot$, and the Bohr radius, $a \sim 10^{-9}\,\textrm{--}10^{-8}\,{\rm pc}$, which is comparable to the radial position of the companion several years before the coalescence.\footnote{See Refs.~\cite{Annulli:2020ilw, Annulli:2020lyc} for the discussions on response of ultralight dark matter soliton to external perturbations.}

The most straightforward way to compute the density profile including the decay is to use Eq.~\eqref{halo_w_decay}, which is shown in Figure~\ref{fig:wave_profile}.
For this figure, we choose $m_1 =10^4M_\odot$, $m_2 = M_\odot$, $\gamma_{\rm sp} =7/3$, and $a = 1 / m \alpha_G = 1/ G m_1 m^2 \in [ 10^{-12}, 10^{-9}] \,{\rm pc}$.
As one can see, the core is absent and the inner part of density profile is replaced by a power-law profile of $\rho \sim r^{2\ell_c}$ where the critical angular momentum $\ell_c$ is defined such that $\max_n 2 \Gamma_{n \ell_c} t_{\rm halo} =1$; the modes with $\ell_c$ are the modes with the lowest angular momentum that survive over $t_{\rm halo}$. 

For a later numerical purpose, we provide a simple analytical expression for the density profile. 
Since low angular momentum modes are absorbed, and the density profile begins to have non-vanishing value at $r \gg a$, taking the continuum limit is a good approximation in most cases. 
We approximate the discrete sum to the continuous integral in Eq.~\eqref{halo_w_decay}, and then introduce the lower bound on angular momentum integral in Eq.~\eqref{classical} as $L_c = \ell_c ( \ell_c+1)/m^2$, where $\ell_c$ is the solution of 
$$
2 \max_n\Gamma_{n\ell} t_{\rm halo} = 1.
$$
Here the principal quantum number is chosen as $n = \max(\ell+1 , \frac{1}{3}\sqrt{(\ell+1)(\ell+2)(2\ell+3)})$, which is an approximate value maximizing the decay rate for a given angular momentum $\ell$. 
The resulting density profile is
\bea
\rho(r) \approx \rho_{\rm sp} \left( \frac{r_{\rm sp}}{r} \right)^{\gamma_{\rm sp}} 
\left( 1 - \frac{R_c}{2r} \right)^{\gamma_{\rm sp}} 
\label{wave_approx}
\eea
with $R_c = a \ell_c (\ell_c +1)$. 
This approximation reliably reproduces the profile from the discrete summation for $\ell_c \gg 1$. 

\subsection{Dynamical friction}
The dimensionless coefficient $C(v)$ in Eq.~\eqref{Efdot} encodes the microscopic nature of dark matter. 
For particle dark matter, one finds~\cite{2008gady.book.....B, Hui:2016ltb}
\begin{equation} 
	C_{p}(v) = \frac{1+\Lambda}{\Lambda}\log\bigg[ 1+\Lambda + \sqrt{\Lambda (2+\Lambda)}\bigg]- \sqrt{1+\frac{2}{\Lambda}}
\end{equation}
with $\Lambda = v^2 r/ (Gm_2)$.
Note that $\Lambda = b_{\rm max} / b_{90}$ where $b_{\rm max} =r$ is the maximum impact parameter and $b_{90} = Gm_2/v^2$ is the impact parameter at which the encounter results in $90^\circ$ deflection of the trajectory.
In the limit $\Lambda \gg 1$, the above expression reproduces $C_{p} \approx \ln 2\Lambda$.

For the wave dark matter, the $C$-factor becomes~\cite{Hui:2016ltb}
\bea
C_{w}(kr) \approx {\rm cin}(2kr)-1 + \frac{\sin (2kr)}{2kr} 
\eea
where $k  = mv$ is the wavenumber of dark matter. 
This expression is valid when $b_{90}/\lambda_{\rm dB} = G m_2 m /v \ll1$, which is always satisfied in our case.
The cosine integral is defined as ${\rm cin}(z) = \int^z_0 (1-\cos t) dt/t$.

\subsection{Backreaction}
Due to dynamical friction, the orbit of the companion decays faster. 
As a result, the companion injects energy into the halo, and the injected energy could change the density profile of the halo, especially when it is of the order of the gravitational binding energy. 
The implications of this backreaction to the halo were first investigated in the work of Kavanagh et al~\cite{Kavanagh:2020cfn} by numerically solving the kinetic equation. 
It was found that the injected energy can be significant for $q = m_2 / m_1 \gtrsim 10^{-4}$ such that it greatly reduces the importance of dynamical friction. 

To include the backreaction effect without solving the kinetic equation numerically, we instead model the backreaction effect in the following way. 
Suppose the companion object is located at the radial position $r$. 
If the orbital energy loss via dynamical friction is larger than the gravitational binding energy of the halo, this energy  heats the dark matter particles and suppresses the dark matter density, effectively halting the dynamical friction. 
Based on this observation, we assume that the maximum amount of energy that can be dissipated from the companion is limited by the kinetic energy stored in a shell of halo of thickness $\Delta r$ at $r$ , which is roughly half of the corresponding gravitational binding energy;  the maximum energy loss is given by $\min( \Delta E_f , \Delta U/2)$, where $\Delta U$ is the gravitational binding energy of the halo,
$$
\Delta U = - \frac{G [ m_1 + m_{\rm enc}(r) ]}{r} 4\pi r^2 \rho(r) \Delta r
$$
with the enclosed mass $m_{\rm enc} = 4\pi \int_0^r dr' \, r'^2 \rho(r')$. 
In other words, the dynamical friction fore is replaced as
\bea
\frac{dE_f}{dt}
\to \min\Big( \frac{dE_f}{dt}, \frac{1}{2} \frac{dU}{dt} \Big)
\approx \Big( 1/ \dot{E}_f + 2 / \dot{U} \Big)^{-1}
\label{backreaction_model}
\eea
where the second expression represents the practical implementation of the model in our numerical analysis. 
This model of backreaction reproduces the numerical result obtained by Kavanagh et al~\cite{Kavanagh:2020cfn}, which is shown in Appendix~\ref{app:comparison}.

\subsection{Waveform}
We investigate how the dynamical friction from compressed halo affects the waveform of the emitted gravitational waves. 
The detector output without noise is $\tilde h(f) = F_+ \tilde h_+(f) + F_\times \tilde h_\times (f)$, where $F_{+,\times}$ are the detector pattern functions and $\tilde h_{+,\times}$ are the strains for each polarization state. 
For simplicity, we consider the strain averaged over the sky position, polarization angle, and inclination angle. 
The angle-averaged strain is then given by~\cite{Robson:2018ifk}
\bea
\tilde h(f) = \sqrt{\frac{4}{5}} A(f) e^{i\Psi(f)}. 
\eea
The amplitude and the phase are given by
\bea
A(f) &=& \frac{2}{D_L} \frac{(GM_c)^{5/3} (\pi f)^{2/3}}{\dot{f}^{1/2}}  , 
\\
\Psi(f) &=& 2\pi f [ t_0 + t(f) ] - \Phi_0 - \Phi(f) - \pi /4  ,
\eea
where $D_L$ is the luminosity distance of the event, $t_0$ and $\Phi_0$ are some reference time and constant phase factor, and 
\bea
t(f) &=& \int^f df' \, (1/ \dot{f}'),
\\
\Phi(f) &=& 2 \pi \int^f df'  \, ( f' / \dot{f}' ). 
\eea
Since the dynamical friction modifies the gravitational wave frequency evolution, it affects the waveform by changing the amplitude $A(f)$ and the phase $\Psi(f)$. 

The change in the amplitude $A(f)$ turns out to be a small effect. 
If we consider the evolution of the inspiral several years before it coalesces, the orbital energy loss due to the dynamical friction is subdominant to the energy loss due to the gravitational wave emission in most cases.
That is, the evolution $\dot{f}/f$ is dominated by the gravitational wave, and therefore, the amplitude $A(f)$ is determined by the gravitational wave emission to the leading order; the dynamical friction plays a subleading role for $A(f)$. 

\begin{figure}[t]
\centering
\includegraphics[width=0.45\textwidth]{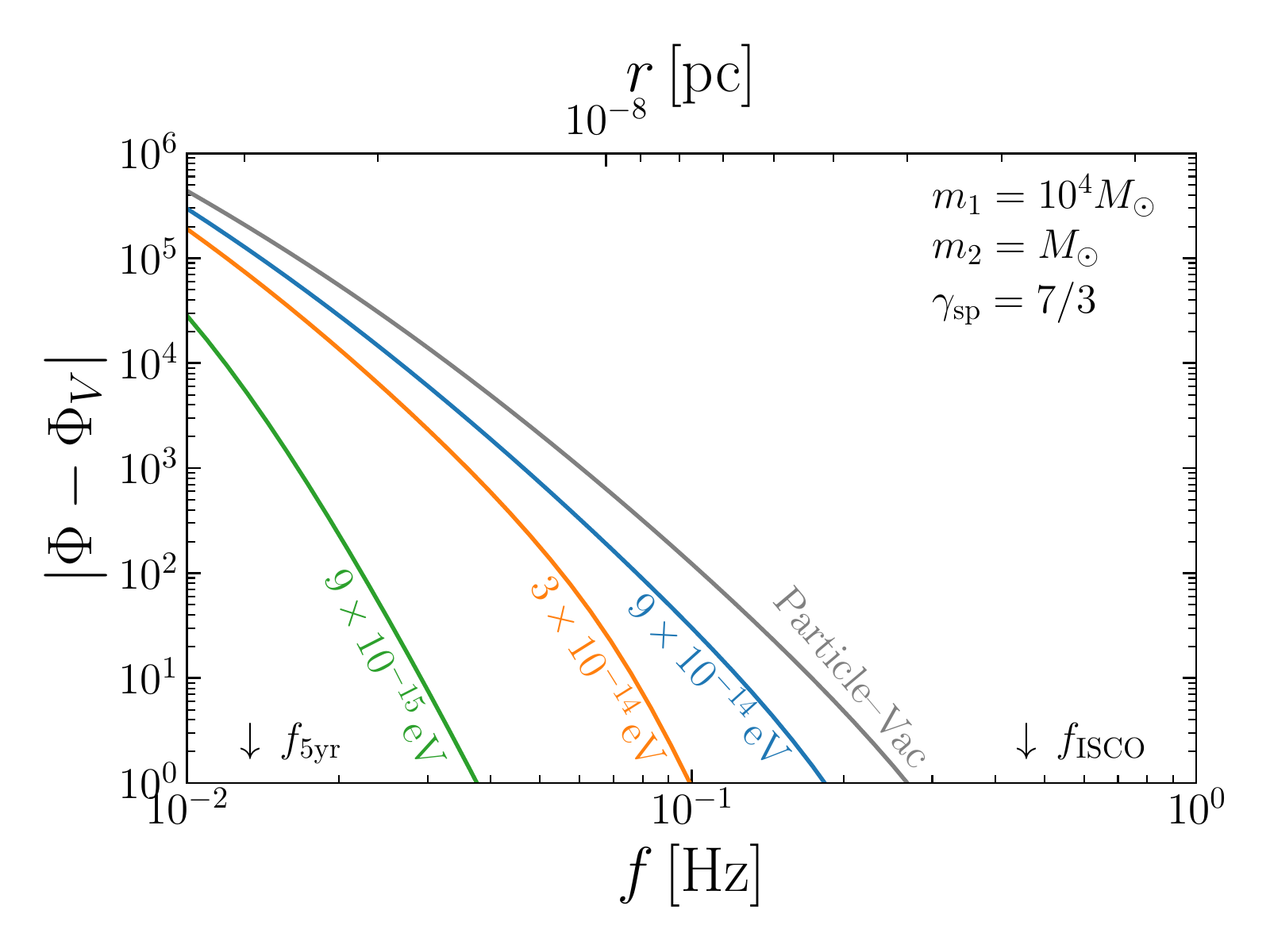}
\includegraphics[width=0.45\textwidth]{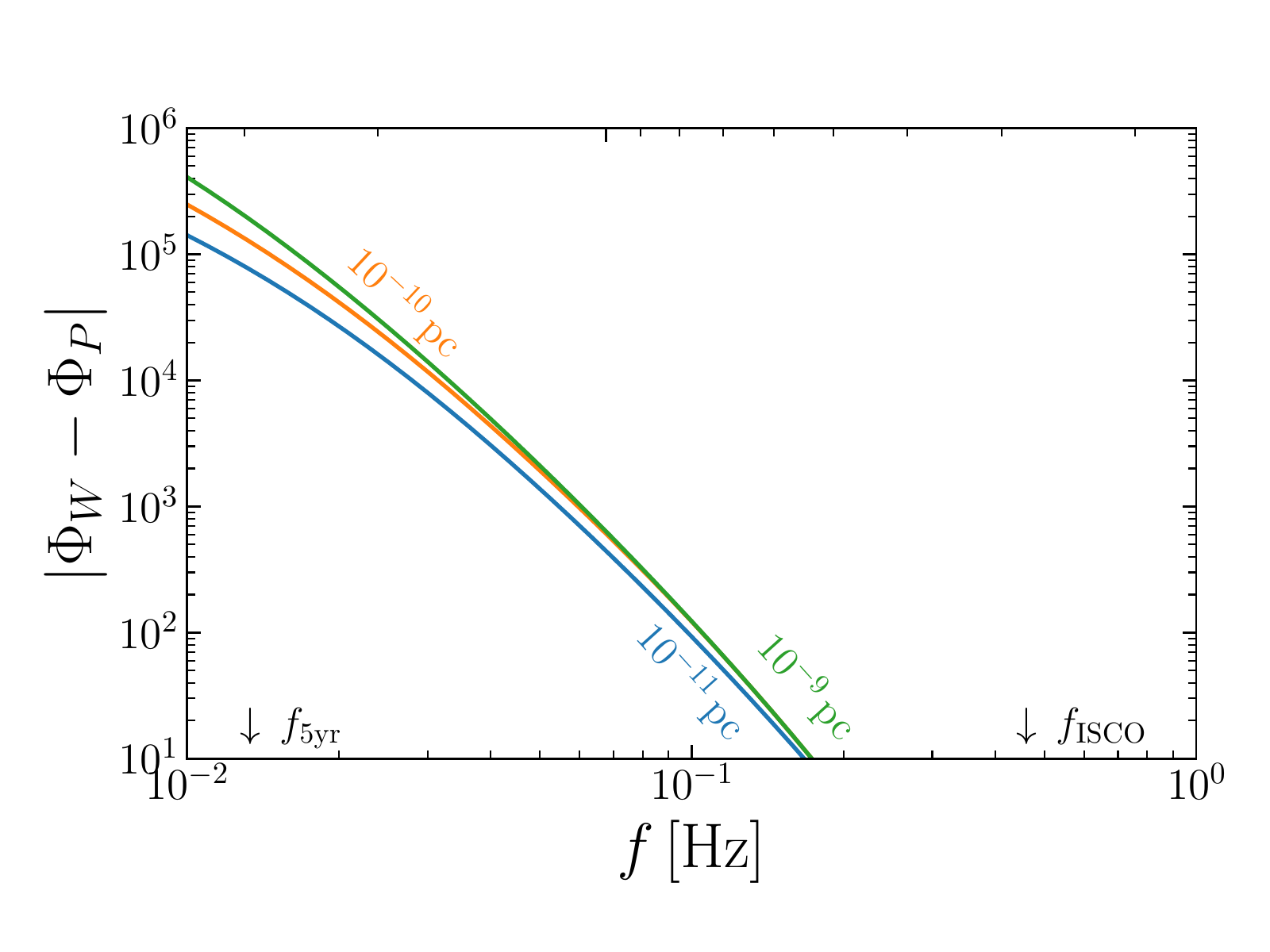}
\caption{(Top) the dephasing of gravitational waves from a compressed dark matter halo with respect to those without a compressed halo. 
The gray line represent the result from the particle dark matter halo.
The blue, orange, and green lines are the results from wave halos with the Bohr radii $a = 10^{-11},\, 10^{-10},\, 10^{-9}$ pc, respectively.
We also show the particle mass for each line. 
(Bottom) the phase difference of gravitational waves from wave dark matter halo and from the particle dark matter halo. 
The color code is the same in both figures. 
Here $f_{5\rm yr}$ denotes the frequency of gravitational waves five years before the coalescence. }
\label{fig:dephasing}
\end{figure}

The phase, however, can be significantly modified by the dynamical friction as it accumulates over time. 
To illustrate this, we compute the phase difference between the gravitational waves from the black hole binary with and without the dynamical friction from the compressed halo. 
In particular, we compute
\bea
\Delta \Phi(f) 
=  2 \pi \int^{f_{\rm ISCO}}_f df'  \, \bigg[( f' / \dot{f}' ) -  ( f' / \dot{f}' )_V\bigg]
\eea
where $(\dot f / f)_V = (96/5) ( \pi f)^{8/3} ( G M_c )^{5/3}$ is the gravitational wave frequency evolution without the dynamical friction. 
In the top panel of Figure~\ref{fig:dephasing}, we show this phase difference, or dephasing, for the particle halo and wave halo with respect to the vacuum case for which there is no compressed halo at all.
We see that the phase difference accumulated over several years could be many orders of magnitude larger than unity. 

In the bottom panel of Figure~\ref{fig:dephasing}, we also show the phase difference of gravitational waves from the wave halo with respect to the particle halo. 
Due to the difference in density profile and the Coulomb logarithm factor, the phase difference is large for the chosen Bohr radii.

\subsection{Parameter estimation}
A natural question is if the phase difference allows us to reconstruct the compressed wave halo parameters, such as the Bohr radius or the wave dark matter mass.
To answer this question, we perform a parameter estimation analysis with the sensitivity of LISA. 
There are 5 parameters related to the system:
\bea
\theta = \{ M_c ,q , \rho_6 , \gamma_{\rm sp}, a \}
\eea
where $M_c = (m_1 m_2)^{3/5}/(m_1+m_2)^{1/5}$ is the chirp mass, $q = m_2/m_1$, $\gamma_{\rm sp}$ is the index of the compressed halo profile, and $a$ is the gravitational Bohr radius. 
Instead of $\rho_{\rm sp}$, we use the density at $r_6 = 10^{-6}\,{\rm pc}$, and $\rho_6 = \rho_{\rm sp} ( r_{\rm sp}/ r_6)^{\gamma_{\rm sp}}$ following~\cite{Coogan:2021uqv}.
There are three more parameters in the waveform, $\theta_c = \{ D_L , t_0 ,\Phi_0\}$, but for the parameter estimation, we choose maximum likelihood values for these parameters for a given $\theta$ to reduce the computational burden.

The detector output is $d(t)=h(t)+n(t)$ where $h(t)$ is the gravitational wave and $n(t)$ is Gaussian detector noise.
The likelihood function is  
\begin{equation}
	{\cal L}(\theta, \theta_c) = {\cal N} \exp\bigg[ -\frac{ 1 }{2} ( d-h | d-h )  \bigg] ,
\end{equation}
where ${\cal N}$ is the normalization constant. 
The inner product $ (a | b) $ is defined as
\begin{equation}
	( a | b ) = {\rm Re}   \int_{-\infty}^{\infty} df\frac{a^* (f) b(f)}{\frac{1}{2} S_n(f)} 
\end{equation}
Here $ S_n(f) $ is the noise power spectral density defined as $\langle n(f) n^*(f') \rangle = \frac{1}{2} \delta(f-f') S_n(f)$. 
For the LISA sensitivity, we use $S_n(f)$ provided in Robson et al~\cite{Robson:2018ifk}.

After fixing $\theta_c$ to the maximum likelihood values, 
the log-likelihood becomes~\cite{Owen:1995tm}
\bea
\!\!\! \ln {\cal L}(\theta) 
= 
\frac{2 \max_{t_0}  \Big| \int_0^\infty df \, e^{2\pi i f t_0} \tilde h^*(f) d(f) /S_n(f) \Big|^2 }{\int_0^\infty df\,  |\tilde h(f)|^2 / S_n(f) }. 
\eea
Here the waveform should be understood as $\tilde h(f,\theta)  = \tilde h(f;\theta)|_{\Phi_0 = 0 , \, t_0 = 0}$. 
The quantity in the numerator is simply the maximum of the Fourier transformation of $\tilde h^*(f) d(f) / S_n(f)$. 
For the parameter estimation, we inject the signal according to the compressed wave dark matter halo, while ignoring the detector noise, i.e. $d = h(\theta_{\rm true})$ with $\theta_{\rm true}$ given by the benchmark values in Table~\ref{tab:bench_posterior}. 
The lower frequency for the above integral is chosen to be the frequency of GWs five years before the coalescence, and the upper frequency is chosen to be $f_{\rm upper} = \min(1\,{\rm Hz}, f_{\rm ISCO})$ where $f_{\rm ISCO}$ is the frequency of gravitational waves at the inner most stable circular orbit $r_{\rm ISCO} = 6Gm_1$.

\subsection{Result}
For the parameter estimation, we choose the following benchmark for $\theta_{\rm true}$:
\begin{table}[h]
\begin{tabular}{|c|c|c|c|c|c|c|}
\hline
& $M_c \,[M_\odot]$ 
& $q$ & $\rho_6\, [10^{15}M_\odot/{\rm pc}^3]$ 
& $\gamma_{\rm sp}$ 
& $a$ [pc] 
& $D_L$ [Mpc]
\\ \hline
$\theta_{\rm true}$
& $ 39.8 $ 
& $10^{-4}$ 
& $25$ 
& $7/3$ 
& $ 10^{-11}$
& 203
\\ \hline
\end{tabular}
\caption{Benchmark values for the wave halo for the parameter estimation.}
\label{tab:bench_posterior}
\end{table}\\
In terms of black hole masses, the above corresponds to $m_1 =10^4M_\odot$ and $m_2 =M_\odot$. 
The density $\rho_6$ is obtained from the benchmark halo in Eq.~\eqref{benchmark_halo} with the central black hole $m_1 =10^4M_\odot$. 
We assume that LISA measures the last five years of inspiral before the coalescence. 
For the numerical computation of the posterior distribution, we use the publicly available nested sampler \texttt{dynesty}~\cite{Speagle:2019ivv}. 
The above benchmark corresponds to a signal-to-noise ratio ${\rm SNR} \simeq 15$. 

\begin{figure*}[t]
\centering
\includegraphics[width=0.9\textwidth]{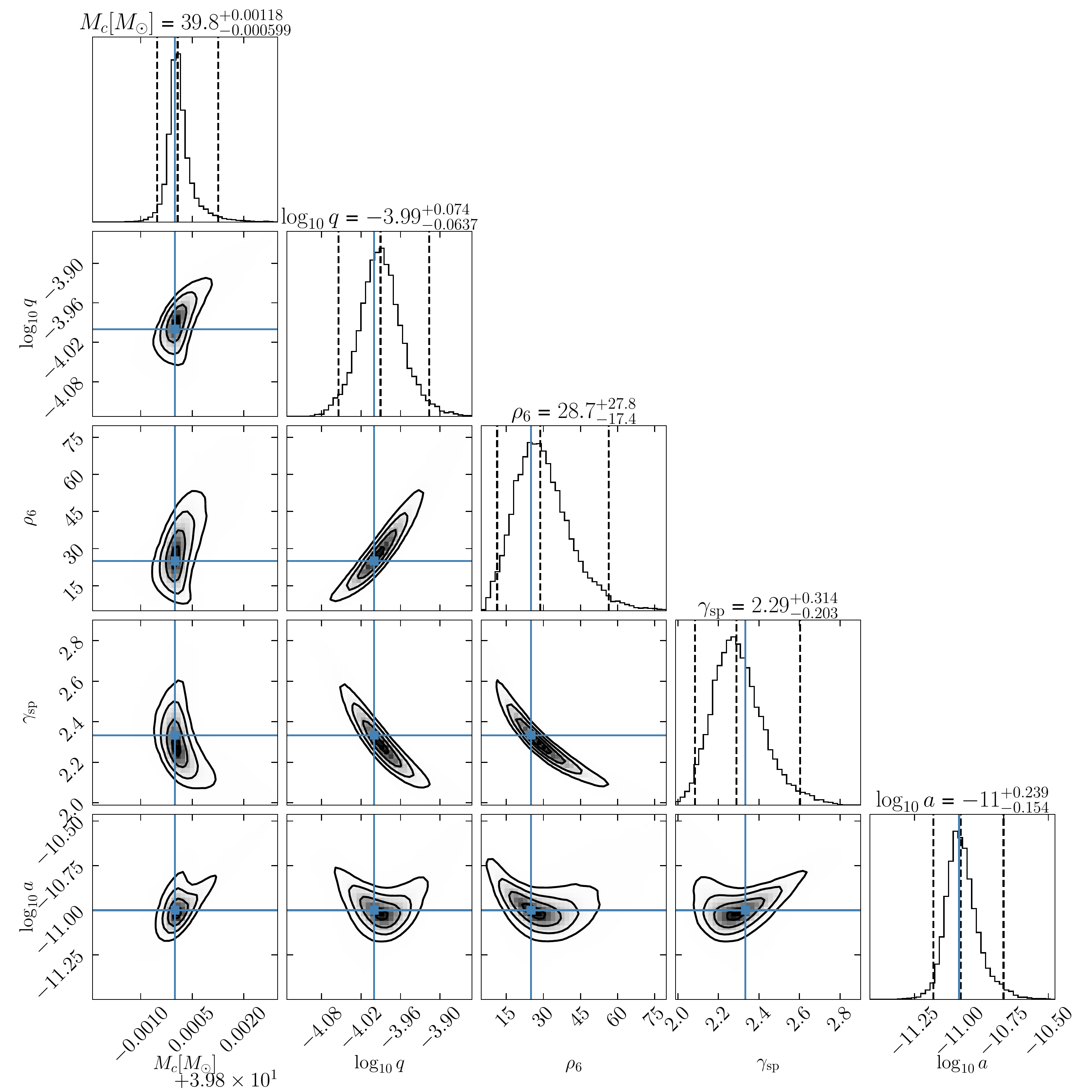}
\caption{Posterior distribution for an intermediate-mass ratio inspiral within the compressed wave halo.
The signal-to-noise ratio is ${\rm S/N}  \simeq 15$.  
Quoted numbers on the top of 1D histogram represents $2.5\%$ and $97.5\%$ quantiles; the contours in 2D histogram represents $0.5,\,1,\,1.5,\, 2\sigma$ levels. 
The compressed wave halo can be reconstructed with five years of gravitational wave observations before its coalescence.
We choose $a = 10^{-11} \,{\rm pc}$ in this example, which corresponds to  the particle mass $m \simeq 10^{-13}\eV$. }
\label{fig:wave_posterior}
\end{figure*}

In Figure~~\ref{fig:wave_posterior}, we show 1D and 2D marginalized posterior distribution for the wave dark halo for the above benchmark scenario. 
Some of the parameters, such as the chirp mass $M_c$, the mass ratio $q$, and the Bohr radius $a$, have resolved peaks around the true value with relatively small errors as one can see in their 1D marginalized distribution.
On the other hand, the posterior distribution for the above benchmark does not provide strong information on the other two parameters, $\rho_6$ and $\gamma_{\rm sp}$. 
We also observe strong degeneracy among parameters, for instance, between $\rho_6$ and $\gamma_{\rm sp}$, and between $\rho_6$ and $\log_{10}q$. 
The chosen gravitational Bohr radius corresponds to $m \simeq 10^{-13}\eV$.

We also compute the Bayes factor to see if the gravitational waves from wave halo can be distinguished from those of a particle halo. 
The Bayes factor is defined as
\bea
B(d)  = \frac{Z_{w} (d) }{Z_p (d)}
\label{bayes}
\eea
where the evidence $Z_i$ is given
\bea
Z_i  = \int d\theta \,  {\cal L}_i(\theta) \pi_i(\theta),
\eea
with $i = w,p$ representing the wave and the particle halo model, respectively. 
To compute the evidence, we inject the signal according to a wave dark matter halo, and compute the evidence with the wave halo model and also with the particle halo model. 
For the Bayes factor computation, we choose three benchmarks shown in Table~\ref{tab:bench}. 
\begin{table}[h]
\begin{tabular}{|c|c|c|c|c|c|}
\hline
& $M_c \,[M_\odot]$ & $q$ & $\rho_6\, [10^{15}M_\odot/{\rm pc}^3]$ & $\gamma_{\rm sp}$  & $D_L$ [Mpc]
\\ \hline
B1 & $ 22.2 $ & $10^{-3}$ & $6.8$ & $7/3$ & 83 
\\ \hline
B2 & $ 39.8 $ & $10^{-4}$ & $25$ & $7/3$ & 203
\\ \hline
B3 & $ 10^2 $ & $10^{-5}$ & $120$ & $7/3$ & 750
\\ \hline
\end{tabular}
\caption{Benchmark scenarios for the evidence computation.}
\label{tab:bench}
\end{table}
Each benchmark corresponds to ${\rm SNR}\simeq 15$. 
B1 corresponds to $m_1 =1400M_\odot$ and $m_2 =1.4M_\odot$, while B3 corresponds to $m_1 = 10^5M_\odot$ and $m_2 =M_\odot$. 
B2 is the same as the one considered for the posterior distribution. 
We compute the Bayes factor while varying the gravitational Bohr radius for $a \in [ 10^{-12}, 2\times 10^{-10}] \,{\rm pc}$.
In each case, the prior range is chosen as $M_c/M_{\odot} = M_{c,\rm true}/{M_\odot} \pm 10^{-2}$, $\log_{10} q = \log_{10} q_{\rm true} \pm 0.5$, $\gamma_{\rm sp} \in [ 2.25, 2.5]$, and $\log_{10} a/{\rm pc} \in [-13,-9]$.  
For $\rho_6$, we choose $\rho_6/ (10^{15} M_\odot/{\rm pc}^3 ) \in [0,20], \, [0, 100], \, [0,500]$ for B1, B2, B3, respectively.

\begin{figure}[t]
\centering
\includegraphics[width=0.45\textwidth]{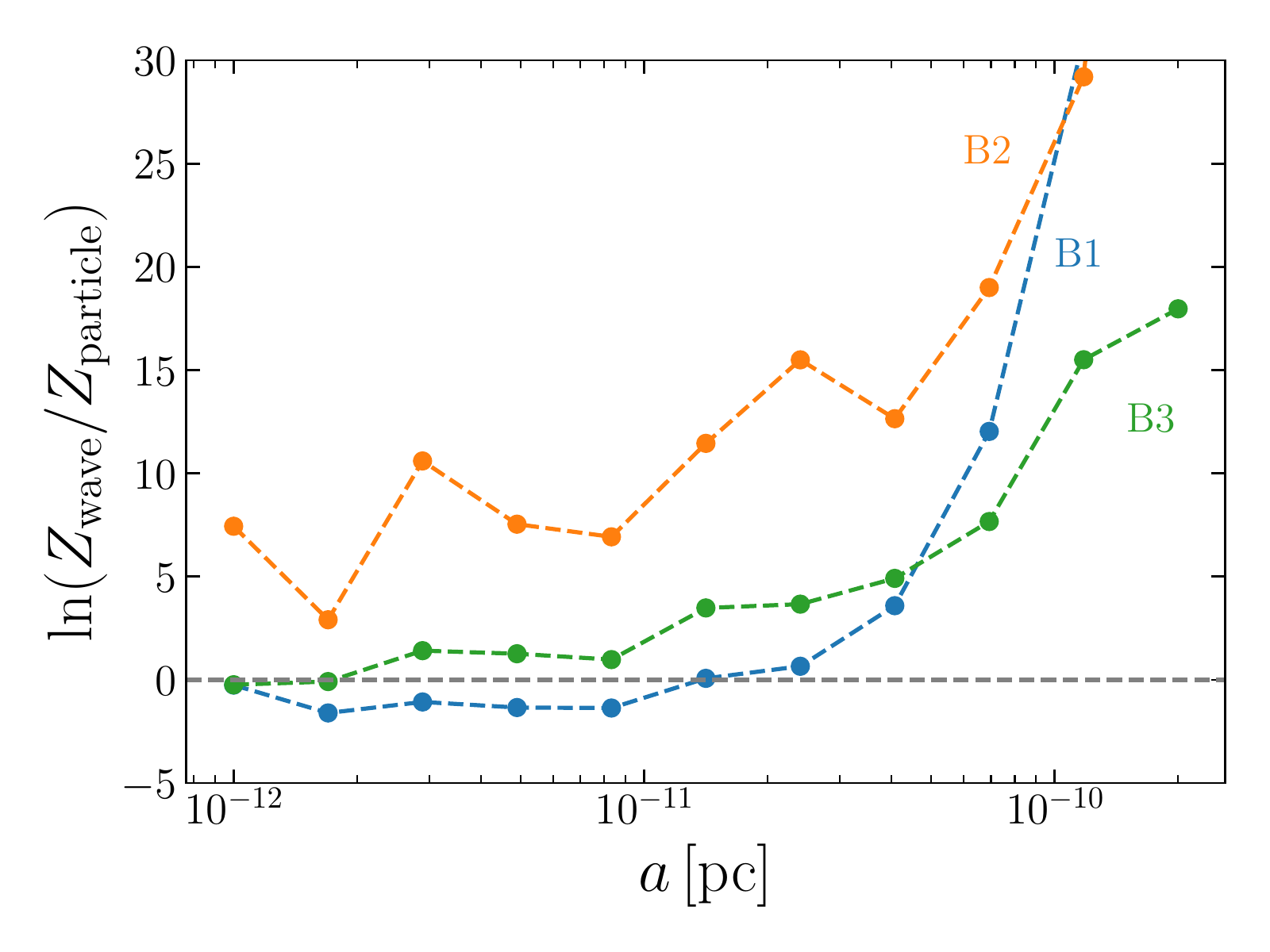}
\caption{The difference of the log-evidence between the wave and the particle halo models, when the input is generated according to a black hole binary in the compressed wave halo. 
We consider the benchmark scenarios listed in Table~\ref{tab:bench}.}
\label{fig:bayes_wave}
\end{figure}

The Bayes factor is shown in Figure~\ref{fig:bayes_wave}. 
In all cases, the log-Bayes factor is close to zero for sufficiently small Bohr radius, meaning that the particle halo is able to fit the data even though the data is generated according to the wave model.
This is because the wave density profile converges to that of particle halo for small Bohr radius.
While there is always some difference induced by the Coulomb logarithm, such difference is not sufficient to distinguish the wave halo from the particle halo. 
As the gravitational Bohr radius increases, the log-Bayes factor increases, indicating that the wave model is preferred over the particle model. 
Especially for B2, the log-Bayes factor increases to ${\cal O}(1)$ values at relatively smaller Bohr radius, $a \simeq {\cal O}(1) \times 10^{-12}\,{\rm pc}$, providing the most promising physics scenarios to identify the wave nature of dark matter from gravitational wave observations.

\section{Discussion}\label{sec:discussion}
We discuss several assumptions and approximations that we have made in the previous sections and investigate how they affect our results. 

\subsection{Accretion}
We have assumed that the companion mass $m_2$ is constant.
If the companion is a black hole, it may accrete masses from the surrounding dark matter halo. 
This would affect the dynamical friction ($dE_{f}/dt\propto m_2^2$) as well as the gravitational wave emission ($dE_{\rm GW}/dt\propto m_2^2$).

The mass accretion rate is
\bea
\frac{dm_2}{dt} = \rho (\sigma_{\rm abs} v),
\eea
where the absorption cross section $\sigma_{\rm abs}$ is computed by Unruh~\cite{Unruh:1976fm}. 
For $v \lesssim 2\pi G m_2 m$, $\sigma_{\rm abs} = 32\pi^2 (G m_2)^3 m / v^2$; for $v \gtrsim 2\pi G m_2 m$, $\sigma_{\rm abs} = 16\pi (G m_2)^2/v$. 
The circular velocity for $r <r_{5 \rm yr}$ is larger than $2\pi G m_2 m = 7\times 10^{-4} (m_2 / 1.4M_\odot) (m / 10^{-14}\eV)$, and therefore, the accretion rate can be approximated as $dm_2 / dt \approx 16\pi (Gm_2)^2 \rho$.
We check numerically that the accretion over five years before the coalescence remains at $\Delta m_2/m_2 \sim {\cal O}(10^{-4})$ for all cases. 
More detailed analyses can be found in \cite{Traykova:2021dua,Vicente:2022ivh}, where the dynamical friction and the accretion effect are derived in a consistent framework.

\subsection{Relaxation}
Gravitational interaction among wave dark matter can modify the halo profile over time through the relaxation process.
It is therefore important to check if the relaxation takes place in the system that we considered in the previous section.  

We estimate the relaxation time scale.
In the wave halo, the relaxation can be understood as a result of gravitational interaction between quasiparticles, whose size is the wavelength of dark matter and whose mass is the total mass enclosed within the de Broglie volume~\cite{Hui:2016ltb}. 
The relaxation time scale is defined as the timescale during which these quasiparticles exchange the kinetic energy by an order one factor. 
More specifically, the relaxation time scale can be estimated as~\cite{1987degc.book.....S, 2008gady.book.....B}
\bea
t_{\rm rel}
= \frac{\sigma^2}{ D[ (\Delta v_{||})^2]} \bigg|_{ {v = \sqrt{3}\sigma} }
\simeq 0.08 \frac{m^3\sigma^6}{G^2 \rho^2 \ln \Lambda}
\label{t_relax}
\eea
where $\sigma^2 \approx v_c^2 / (1+\gamma_{\rm sp})$ is the velocity dispersion and the diffusion coefficient $D[(\Delta v_{||})^2]$ is 
\bea
D[( \Delta v_{||})^2] \simeq
13\times \frac{G^2 \rho^2(r) \ln\Lambda}{m^3 \sigma^4} 
\label{Diff2}
\eea
where $\ln\Lambda \simeq \ln m \sigma r$ is the Coulomb logarithm.
The numerical coefficient has a mild dependence on $\gamma_{\rm sp}$; it changes from $13.4$ to $12$ for $\gamma_{\rm sp} \in [2.25,2.5]$. 
After the relaxation time scale, wave modes have exchanged energy by an order one factor, replenishing low angular momentum modes that are subsequently absorbed by the central black hole, suppressing the wave halo density further. 
A detailed derivation of the relaxation time scale on compressed halo is discussed in Appendix~\ref{app:relaxation}. 

\begin{figure}
\centering
\includegraphics[width=0.45\textwidth]{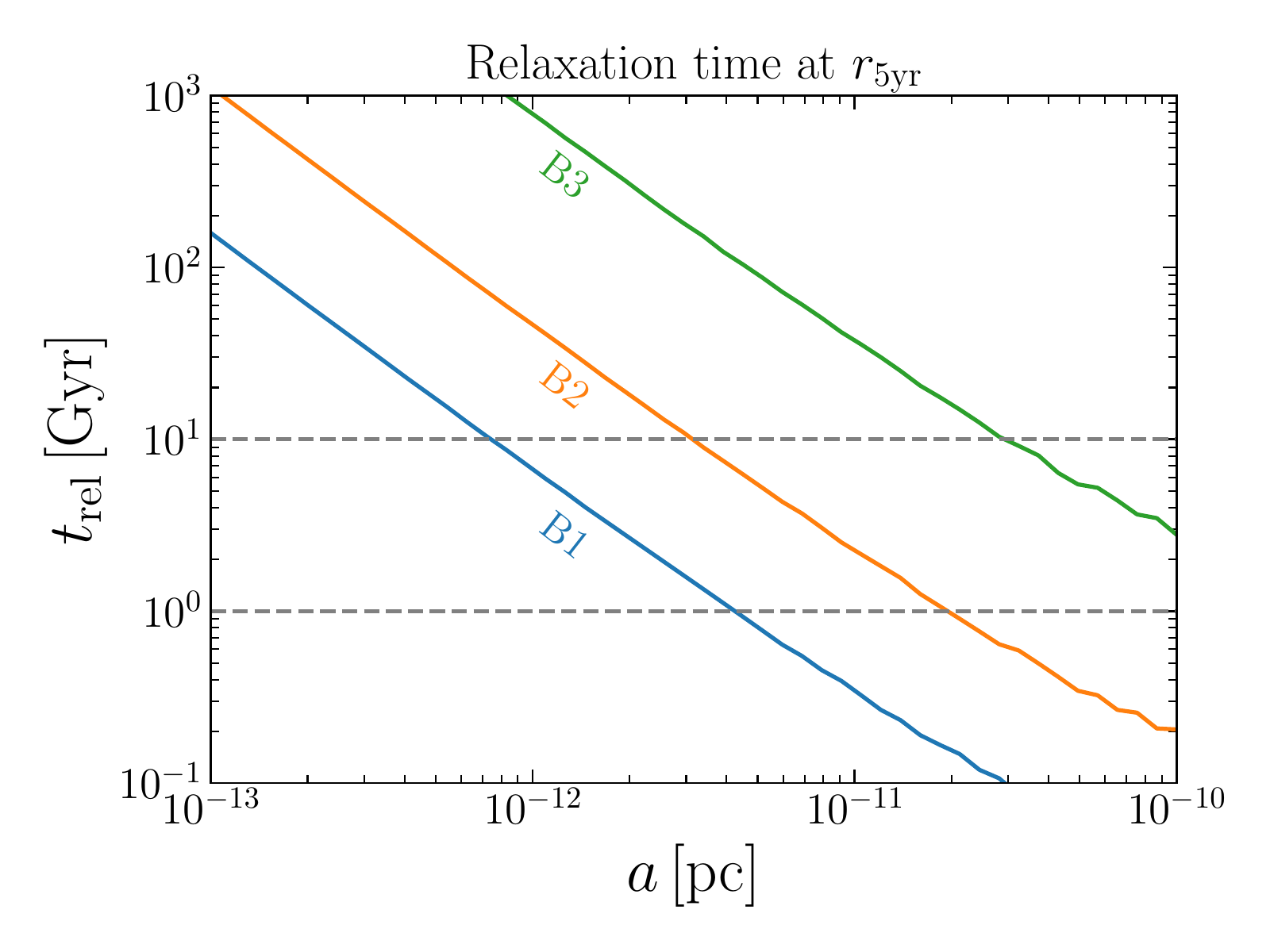}
\caption{The relaxation time scale at $r_{5\rm yr}$ for each benchmark introduced in the previous section. 
As we increase the mass of the central black hole, the relaxation time scale generally increases. 
If the relaxation time scale is significantly smaller than Gyr time scale, it is expected that the wave halo profile is further suppressed as the low angular momentum will be replenished through the relaxation process.  }
\label{fig:relaxation}
\end{figure}

Since the most important contribution to the dephasing arises around $r \sim r_{5\rm yr}$, we compute the relaxation time scale at $r_{\rm 5yr}$ for each benchmark B1, B2, and B3. 
In Figure~\ref{fig:relaxation}, we show the relaxation time scale of the halo as a function of the Bohr radius. 
For B1, the relaxation time scale becomes smaller than Gyr scale for the Bohr radius $a \gtrsim 10^{-11}\,{\rm pc}$ ($m \lesssim 2\times 10^{-13}\eV$). 
In such cases, the wave halo is subject to the relaxation, and hence, the density profile is further suppressed from what we discussed in the previous section.
To correctly model the density profile, one needs to investigate the dynamical evolution of the wave modes due to gravitational interactions, which is beyond the scope of this work.
We also compute the relaxation time scale for the other benchmark scenarios. 
As the mass of central black hole increases, the relaxation time scale also increases at a given Bohr radius. 
In the figure, we see that the relaxation time scale for these two benchmark scenarios can be Gyr time scale even for $a \gtrsim 10^{-11}\,{\rm pc}$, where the log-Bayes between wave and particle halo begins to exceed unity.
Since the relaxation time scale is comparable to the age of halo, we expect that the wave profile is approximately given by the form~\eqref{halo_w_decay}, and does not significantly change over the age of halo. 

\subsection{Stochastic heating}
Numerical simulations of fuzzy dark matter observed granular structures, which represent fluctuations of the gravitational potential~\cite{Schive:2014dra}. 
These fluctuations of the wave DM halo can be described as quasiparticles, whose size is of the order of its de Broglie wavelength and the mass is given by the mass enclosed within the de Broglie volume~\cite{Hui:2016ltb, Bar-Or:2018pxz}. 
Depending on their mass, quasiparticles could impart non-negligible kinetic energy to stellar objects. 
Such stochastic heating effect affects the orbital evolution of the companion object, and therefore, could potentially change the dephasing pattern. 

\begin{figure}
\centering
\includegraphics[width=0.45\textwidth]{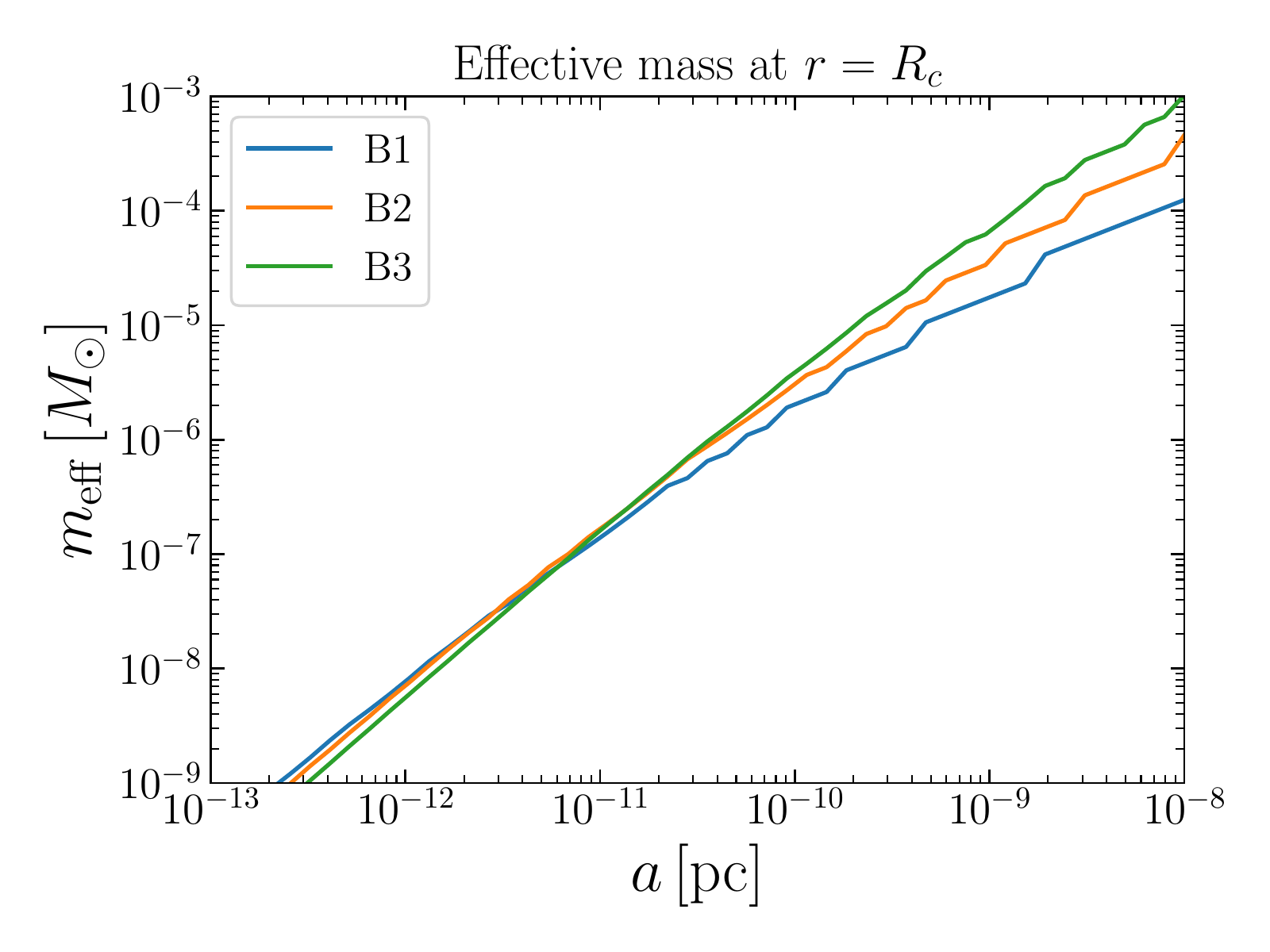}
\caption{The effective mass of quasiparticles for each benchmark. 
For all range of the gravitational Bohr radius, the effective mass is several orders of magnitude smaller than the companion mass. }
\label{fig:eff_dB}
\end{figure}

To estimate the stochastic heating effect, we begin with the change of the energy of the object $m_2$ due to its interaction with the dark matter halo, which can be written as
\bea
\frac{d E_f}{dt } = D[ \Delta E].
\eea
The energy diffusion coefficient $D[\Delta E]$ denotes the change in energy per unit time.
It can be rewritten in terms of velocity diffusion coefficients by noticing $\Delta E /m_2= \frac{1}{2} ( v + \Delta v) - \frac{1}{2} v^2 =  v \cdot \Delta v + \frac{1}{2} (\Delta v)^2$.
This leads to
\bea
D[\Delta E]  = m_2(  v \cdot D [\Delta v] + \frac{1}{2} D[ (\Delta v)^2] ),
\eea
where the diffusion coefficients for velocities in the wave limit are computed in~\cite{Bar-Or:2018pxz, Bar-Or:2020tys, Chavanis:2020upb}. 
The first term is the one we used to compute the dephasing of the gravitational wave, i.e. Eq.~\eqref{Efdot}; the second term is due to the stochastic heating with quasiparticles. 
In our analysis, we ignored the second term. 

The relative importance of the second term is
\bea
\frac{{\rm cooling}}{{\rm heating}}
\approx \frac{ \dot{E}_f }{ D[(\Delta v_{||})^2] }
\sim \frac{m_2}{m_{\rm eff}}
\eea
where $m_{\rm eff} = \rho(r) ( \sqrt{2\pi} / mv_c)^3$ is roughly the mass enclosed within de Broglie volume, which can be interpreted as the mass of quasiparticles. 
The numerator $\dot{E}_f$ is the usual dynamical friction force~\eqref{Efdot}, and the denominator is the second velocity diffusion coefficient, which is already obtained in the previous section, Eq.~\eqref{Diff2}. 
A detailed derivation of the diffusion coefficient is given in Appendix~\ref{app:relaxation}. 
We see that the stochastic heating can be neglected as long as the quasiparticle mass is smaller than the companion mass. 

As it is clear from the expression, the quasiparticle mass increases as $r$ decreases, indicating that the stochastic heating would be the largest at the smallest $r$. 
In Figure~\ref{fig:eff_dB}, we compute the mass of quasiparticle at the radius $r = R_c = a \ell_c ( \ell_c+1)$. 
We see that the stochastic heating is negligible to the dynamical friction in all the benchmarks we consider in this work.

\section{Conclusion}

We discussed how the wave dark matter responds to the adiabatic growth of the black hole. 
Using the adiabatic theorem, we showed that the wave halo is compressed in a similar fashion as the particle halo is compressed in the semiclassical limit. 
The difference arises near the central region, where the compressed wave halo may have a solitonic core.
The existence of the solitonic core depends on the system, as it can be completely swallowed by the central black hole. 
We considered one example where the central soliton in the compressed halo survives over the astrophysical time scale, and one example where it does not. 
If the soliton and other low angular momentum modes are absorbed, the wave density profile behaves as a broken-power law, where it follows the usual spike profile at large radii, $\rho \propto r^{-\gamma_{\rm sp}}$, while at small radii, it behaves $\rho \propto r^{2\ell_c}$, where $\ell_c$ is the lowest angular momentum that can survive for the age of the halo $t_{\rm halo}$. 

Having discussed the adiabatic compression of wave dark matter, we  considered one interesting astrophysical application regarding gravitational waves from intermediate mass-ratio inspirals.
In the presence of compressed wave dark matter, the inspiral certainly experiences additional energy dissipation due to the dynamical friction. 
This additional dissipation of orbital energy causes a dephasing of gravitational wave signals.
Due to the difference in the mass density and the Coulomb logarithm factor, the companion object experiences a dynamical friction force from the wave halo that is distinctive from the force that it would experience in the particle halo. 
We showed that the wave halo can be reconstructed by the gravitaitonal wave observations with the sensitivity given by future LISA mission for certain benchmark values, and that it can also be distinguished from particle halo. 
This provides an interesting way to probe the microscopical nature of the dark matter.
Although we only considered wave halo in this work, it could be interesting to investigate other types of particle dark matter such as self-interacting dark matter or degenerate fermionic dark matter as they can form cored profile and predict distinctive Coulomb logarithm factor for the dynamical friction~\cite{Bar:2021jff}.

\acknowledgements We would like to thank Kfir Blum, Vitor Cardoso, Aleksandr Chatrchyan, Yifan Chen, Joshua Eby, Cem Er{\"o}ncel, Robin Diedrichs, Oindrila Ghosh, Yann Gouttenoire, Eric Madge, Andrea Mitridate, Wolfram Ratzinger, Henrique Rubira, Laura Sagunski, G\"{u}nter Sigl, Tomer Volansky,  and Wei Xue for the useful conversations. 
This work is supported by the Deutsche Forschungsgemeinschaft under Germany’s Excellence Strategy - EXC 2121 Quantum Universe - 390833306. HK and XX would like to express special thanks to the Mainz Institute for Theoretical Physics (MITP) of the Cluster of Excellence PRISMA*(Project ID 39083149) for its hospitality and support.

\appendix

\section{Derivation of relaxation time scale}\label{app:relaxation}
We derive the relaxation time scale Eq.~\eqref{t_relax} in this appendix. 

Before we compute the diffusion coefficient, let us consider the phase space distribution, which allows us to compute the diffusion coefficient as well as the velocity dispersion of the compressed halo. 
The phase space distribution of the compressed halo is determined by the initial phase space distribution as in Eq.~\eqref{particle_phase}
$$
f(E_1, L_1) = f(E_0 (E_1, L_1) )
$$
where $L_0 = L_1 = L$ and the relation between $E_1$ and $E_0$ is determined by the radial action. 
The final phase space distribution in this way depends on the angular momentum. 
Instead of using this distribution, we approximate the final distribution obtained from Eddington's formula Eq.~\eqref{eddington} with $\rho(r) = \rho_{\rm sp} (r_{\rm sp}/r)^{\gamma_{\rm sp}}$. 
In this case, we find
\bea
f(E) = \frac{\rho(r)}{(-2\pi \Phi)^{3/2} }
\frac{\Gamma(\gamma_{\rm sp}+1)}{\Gamma(\gamma_{\rm sp} - 1/2)}
\left( \frac{\Phi}{E} \right)^{3/2 - \gamma_{\rm sp}}. 
\eea
This approximate distribution fully reproduces the density profile, and provide a reasonable approximation for the diffusion coefficient computation. 

Having found the approximate phase space distribution of the compressed halo, we can compute the velocity dispersion as
\bea
\sigma^2 =
\frac{1}{3\rho}
\int d^3v \, v^2 f(v)
= \frac{v_c^2}{1+\gamma_{\rm sp}}.
\eea
Here $v_c^2 = Gm_1/r$ is the circular velocity. 

The diffusion coefficient for the wave dark matter is~\cite{Bar-Or:2020tys, Chavanis:2020upb}
\bea
D[ (\Delta v_{||})^2] &=& 
\frac{32\pi^2 G^2 \ln\Lambda}{3} \frac{(2\pi)^3}{m^3}
\nonumber\\
&&\times\bigg[ \frac{1}{v^3}
\int^{\frac{v^2}{2} + \Phi}_{\Phi} dE \,  [2(E-\Phi)]^{\frac32} f^2(E) 
\nonumber\\
&&\qquad\qquad\qquad
+ \int_{\frac{v^2}{2}+ \Phi}^0 dE\, f^2(E) \bigg]
\label{D2}\\
&=& 
\frac{32\pi^2 G^2\ln\Lambda}{3} 
\frac{\rho^2(r)}{m^3 v_c^4}
\bigg[
\frac{\Gamma(\gamma_{\rm sp}+1)}{\Gamma(\gamma_{\rm sp} - \frac12)} 
\bigg]^2{\cal I}(v/v_c,\gamma_{\rm sp})
\nonumber
\eea
where $\ln\Lambda \simeq \ln (m \sigma r)$ is the Coulomb logarithm in the wave limit and ${\cal I}$ is defined as
\bea
{\cal I}(x,\gamma_{\rm sp})
= \frac{1}{x^3}
\int_{1 - \frac{x^2}{2}}^1 d \eps \frac{[2(1-\eps)]^{3/2}}{\eps^{3-2\gamma_{\rm sp}}}
+ \int_0^{1- \frac{x^2}{2} }
\frac{d\eps}{\eps^{3-2\gamma_{\rm sp}}}  . 
\nonumber
\eea
With this diffusion coefficient, we find
\bea
t_{\rm rel} = \frac{m^3 \sigma^6}{G^2 \rho^2 \ln \Lambda} 
\left[ \frac{3}{32\pi^2}
\left(
\frac{\Gamma(\gamma_{\rm sp} - \frac{1}{2} )}{\Gamma(\gamma_{\rm sp}+1) }
\right)^2
\frac{(1 + \gamma_{\rm sp})^2}{ {\cal I}(v/v_c, \gamma_{\rm sp})}
\right]
\nonumber
\eea
where $ v = \sqrt{3}\sigma$. 
The quantity in the square brackets has a mild dependence on $\gamma_{\rm sp}$; it takes values $[0.075, \, 0.083]$ for $\gamma_{\rm sp} \in [2.25,2.5]$.

\begin{figure}
\centering
\includegraphics[width=0.45\textwidth]{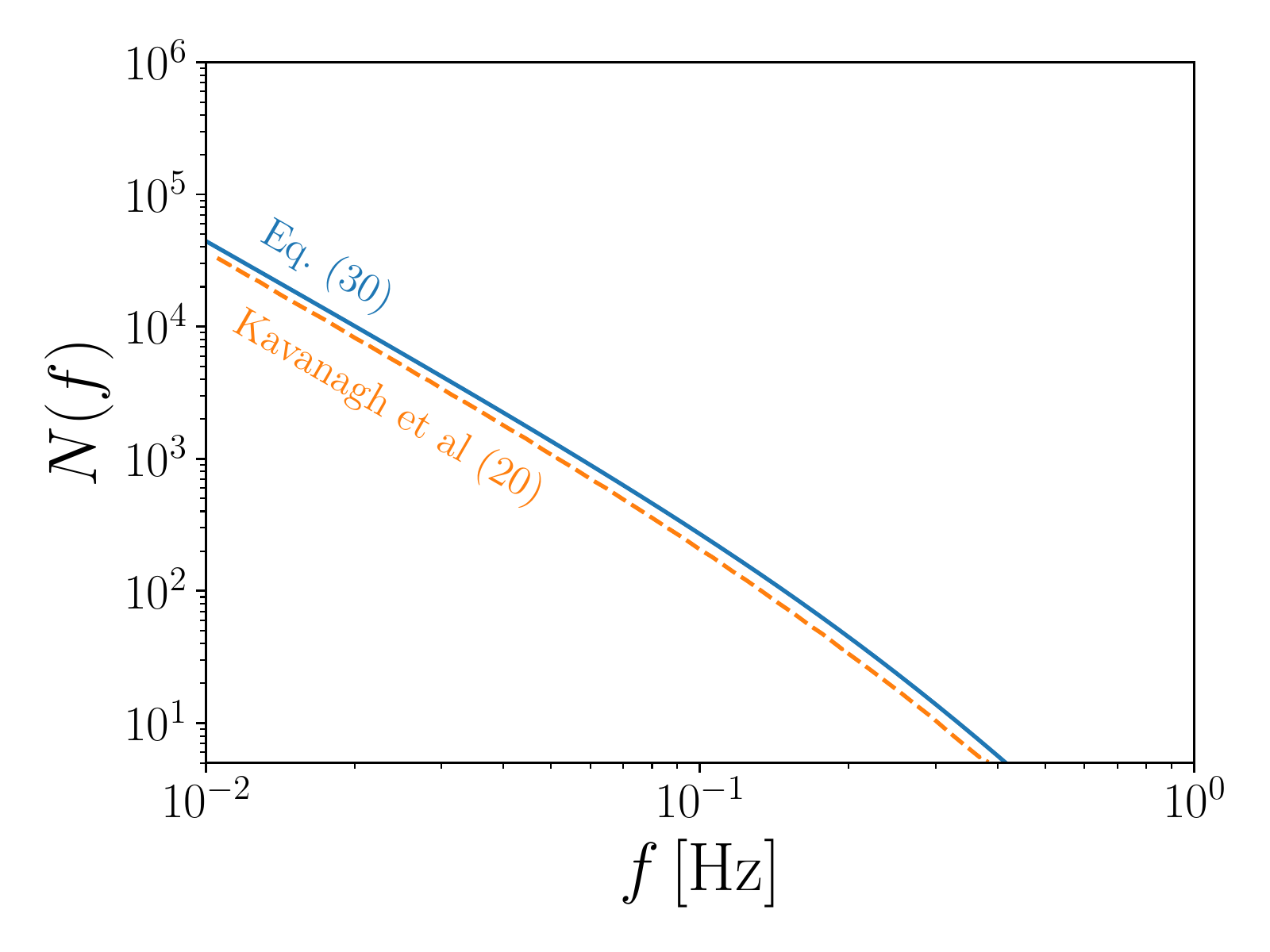}
\caption{The number of cycles of gravitational waves counted from the innermost stable circular orbit. 
The solid line is obtained from our analytical modeling described in the main text, while the dots are numerical results taken from Ref.~\cite{Kavanagh:2020cfn}.
For this figure, we choose $m_1 =1400M_\odot$, $m_2 = 1.4M_\odot$, $\gamma_{\rm sp} = 7/3$, and $\rho_{\rm sp} =226 M_\odot/{\rm pc}^3$. }
\label{fig:comparison}
\end{figure}

\section{Comparison to previous works}\label{app:comparison}
To check if our backreaction model correctly reproduces the numerical result of Kavanagh et al~\cite{Kavanagh:2020cfn}, we compare the number-of-cycles difference in the particle dark matter halo, defined as
\bea
N(f)
&=& \int^{t_{\rm ISCO}}_t dt' \, [f_{p}(t') - f_v(t') ]
\nonumber\\
&=& \int^{f_{\rm ISCO}}_f df' \, [(f' / \dot f')_{p} - (f' / \dot f')_v ]
\eea
where the subscript $p$ and $v$ denotes the evolution of gravitational wave frequencies with and without a particle dark matter halo~\eqref{particle_halo}. 
The result obtained by our analytical model and the numerical result in~\cite{Kavanagh:2020cfn} are shown in Figure~\ref{fig:comparison}. 
Over the relevant frequency range, our backreaction model~\eqref{backreaction_model} overestimates the number of cycle roughly by $20\%$ or less, compared to the numerical result in~\cite{Kavanagh:2020cfn}. 

\bibliographystyle{utphys}
\bibliography{ref}

\providecommand{\href}[2]{#2}\begingroup\raggedright\begin{thebibliography}{10}

\bibitem{1996ApJ...462..563N}
J.~F. {Navarro}, C.~S. {Frenk}, and S.~D.~M. {White}, ``{The Structure of Cold
  Dark Matter Halos},'' \href{http://dx.doi.org/10.1086/177173}{{\em \apj}
  {\bfseries 462} (May, 1996) 563},
  \href{http://arxiv.org/abs/astro-ph/9508025}{{\ttfamily
  arXiv:astro-ph/9508025 [astro-ph]}}.

\bibitem{1997ApJ...490..493N}
J.~F. {Navarro}, C.~S. {Frenk}, and S.~D.~M. {White}, ``{A Universal Density
  Profile from Hierarchical Clustering},''
  \href{http://dx.doi.org/10.1086/304888}{{\em \apj} {\bfseries 490} no.~2,
  (Dec., 1997) 493--508},
  \href{http://arxiv.org/abs/astro-ph/9611107}{{\ttfamily
  arXiv:astro-ph/9611107 [astro-ph]}}.

\bibitem{Salucci:2018hqu}
P.~Salucci, ``{The distribution of dark matter in galaxies},''
  \href{http://dx.doi.org/10.1007/s00159-018-0113-1}{{\em Astron. Astrophys.
  Rev.} {\bfseries 27} no.~1, (2019) 2},
  \href{http://arxiv.org/abs/1811.08843}{{\ttfamily arXiv:1811.08843
  [astro-ph.GA]}}.

\bibitem{1980ApJ...242.1232Y}
P.~{Young}, ``{Numerical models of star clusters with a central black hole. I -
  Adiabatic models.},'' \href{http://dx.doi.org/10.1086/158553}{{\em \apj}
  {\bfseries 242} (Dec., 1980) 1232--1237}.

\bibitem{1986ApJ...301...27B}
G.~R. {Blumenthal}, S.~M. {Faber}, R.~{Flores}, and J.~R. {Primack},
  ``{Contraction of Dark Matter Galactic Halos Due to Baryonic Infall},''
  \href{http://dx.doi.org/10.1086/163867}{{\em \apj} {\bfseries 301} (Feb.,
  1986) 27}.

\bibitem{1995ApJ...440..554Q}
G.~D. {Quinlan}, L.~{Hernquist}, and S.~{Sigurdsson}, ``{Models of Galaxies
  with Central Black Holes: Adiabatic Growth in Spherical Galaxies},''
  \href{http://dx.doi.org/10.1086/175295}{{\em \apj} {\bfseries 440} (Feb.,
  1995) 554}, \href{http://arxiv.org/abs/astro-ph/9407005}{{\ttfamily
  arXiv:astro-ph/9407005 [astro-ph]}}.

\bibitem{1999AJ....117..744V}
R.~P. {van der Marel}, ``{The Black Hole Mass Distribution in Early-Type
  Galaxies: Cusps in Hubble Space Telescope Photometry Interpreted through
  Adiabatic Black Hole Growth},'' \href{http://dx.doi.org/10.1086/300730}{{\em
  \aj} {\bfseries 117} no.~2, (Feb., 1999) 744--763},
  \href{http://arxiv.org/abs/astro-ph/9806365}{{\ttfamily
  arXiv:astro-ph/9806365 [astro-ph]}}.

\bibitem{Gondolo:1999ef}
P.~Gondolo and J.~Silk, ``{Dark matter annihilation at the galactic center},''
  \href{http://dx.doi.org/10.1103/PhysRevLett.83.1719}{{\em Phys. Rev. Lett.}
  {\bfseries 83} (1999) 1719--1722},
  \href{http://arxiv.org/abs/astro-ph/9906391}{{\ttfamily
  arXiv:astro-ph/9906391}}.

\bibitem{Eda:2013gg}
K.~Eda, Y.~Itoh, S.~Kuroyanagi, and J.~Silk, ``{New Probe of Dark-Matter
  Properties: Gravitational Waves from an Intermediate-Mass Black Hole Embedded
  in a Dark-Matter Minispike},''
  \href{http://dx.doi.org/10.1103/PhysRevLett.110.221101}{{\em Phys. Rev.
  Lett.} {\bfseries 110} no.~22, (2013) 221101},
  \href{http://arxiv.org/abs/1301.5971}{{\ttfamily arXiv:1301.5971 [gr-qc]}}.

\bibitem{Eda:2014kra}
K.~Eda, Y.~Itoh, S.~Kuroyanagi, and J.~Silk, ``{Gravitational waves as a probe
  of dark matter minispikes},''
  \href{http://dx.doi.org/10.1103/PhysRevD.91.044045}{{\em Phys. Rev. D}
  {\bfseries 91} no.~4, (2015) 044045},
  \href{http://arxiv.org/abs/1408.3534}{{\ttfamily arXiv:1408.3534 [gr-qc]}}.

\bibitem{Kavanagh:2020cfn}
B.~J. Kavanagh, D.~A. Nichols, G.~Bertone, and D.~Gaggero, ``{Detecting dark
  matter around black holes with gravitational waves: Effects of dark-matter
  dynamics on the gravitational waveform},''
  \href{http://dx.doi.org/10.1103/PhysRevD.102.083006}{{\em Phys. Rev. D}
  {\bfseries 102} no.~8, (2020) 083006},
  \href{http://arxiv.org/abs/2002.12811}{{\ttfamily arXiv:2002.12811 [gr-qc]}}.

\bibitem{Coogan:2021uqv}
A.~Coogan, G.~Bertone, D.~Gaggero, B.~J. Kavanagh, and D.~A. Nichols,
  ``{Measuring the dark matter environments of black hole binaries with
  gravitational waves},''
  \href{http://dx.doi.org/10.1103/PhysRevD.105.043009}{{\em Phys. Rev. D}
  {\bfseries 105} no.~4, (2022) 043009},
  \href{http://arxiv.org/abs/2108.04154}{{\ttfamily arXiv:2108.04154 [gr-qc]}}.

\bibitem{Cole:2022fir}
P.~S. Cole, G.~Bertone, A.~Coogan, D.~Gaggero, T.~Karydas, B.~J. Kavanagh,
  T.~F.~M. Spieksma, and G.~M. Tomaselli, ``{Disks, spikes, and clouds:
  distinguishing environmental effects on BBH gravitational waveforms},''
  \href{http://arxiv.org/abs/2211.01362}{{\ttfamily arXiv:2211.01362 [gr-qc]}}.

\bibitem{Becker:2022wlo}
N.~Becker and L.~Sagunski, ``{Comparing Accretion Disks and Dark Matter Spikes
  in Intermediate Mass Ratio Inspirals},''
  \href{http://arxiv.org/abs/2211.05145}{{\ttfamily arXiv:2211.05145 [gr-qc]}}.

\bibitem{Cline:2022qld}
J.~M. Cline, S.~Gao, F.~Guo, Z.~Lin, S.~Liu, M.~Puel, P.~Todd, and T.~Xiao,
  ``{Blazar constraints on neutrino-dark matter scattering},''
  \href{http://arxiv.org/abs/2209.02713}{{\ttfamily arXiv:2209.02713
  [hep-ph]}}.

\bibitem{Ferrer:2022kei}
F.~Ferrer, G.~Herrera, and A.~Ibarra, ``{New constraints on the dark
  matter-neutrino and dark matter-photon scattering cross sections from TXS
  0506+056},'' \href{http://arxiv.org/abs/2209.06339}{{\ttfamily
  arXiv:2209.06339 [hep-ph]}}.

\bibitem{Chan:2022gqd}
M.~H. Chan and C.~M. Lee, ``{Indirect Evidence for Dark Matter Density Spikes
  around Stellar-Mass Black Holes},''
  \href{http://arxiv.org/abs/2212.05664}{{\ttfamily arXiv:2212.05664
  [astro-ph.HE]}}.

\bibitem{Peccei:1977hh}
R.~D. Peccei and H.~R. Quinn, ``{CP Conservation in the Presence of
  Instantons},'' \href{http://dx.doi.org/10.1103/PhysRevLett.38.1440}{{\em
  Phys. Rev. Lett.} {\bfseries 38} (1977) 1440--1443}.

\bibitem{Weinberg:1977ma}
S.~Weinberg, ``{A New Light Boson?},''
  \href{http://dx.doi.org/10.1103/PhysRevLett.40.223}{{\em Phys. Rev. Lett.}
  {\bfseries 40} (1978) 223--226}.

\bibitem{Wilczek:1977pj}
F.~Wilczek, ``{Problem of Strong $P$ and $T$ Invariance in the Presence of
  Instantons},'' \href{http://dx.doi.org/10.1103/PhysRevLett.40.279}{{\em Phys.
  Rev. Lett.} {\bfseries 40} (1978) 279--282}.

\bibitem{Preskill:1982cy}
J.~Preskill, M.~B. Wise, and F.~Wilczek, ``{Cosmology of the Invisible
  Axion},'' \href{http://dx.doi.org/10.1016/0370-2693(83)90637-8}{{\em Phys.
  Lett. B} {\bfseries 120} (1983) 127--132}.

\bibitem{Abbott:1982af}
L.~F. Abbott and P.~Sikivie, ``{A Cosmological Bound on the Invisible Axion},''
  \href{http://dx.doi.org/10.1016/0370-2693(83)90638-X}{{\em Phys. Lett. B}
  {\bfseries 120} (1983) 133--136}.

\bibitem{Dine:1982ah}
M.~Dine and W.~Fischler, ``{The Not So Harmless Axion},''
  \href{http://dx.doi.org/10.1016/0370-2693(83)90639-1}{{\em Phys. Lett. B}
  {\bfseries 120} (1983) 137--141}.

\bibitem{Graham:2015cka}
P.~W. Graham, D.~E. Kaplan, and S.~Rajendran, ``{Cosmological Relaxation of the
  Electroweak Scale},''
  \href{http://dx.doi.org/10.1103/PhysRevLett.115.221801}{{\em Phys. Rev.
  Lett.} {\bfseries 115} no.~22, (2015) 221801},
  \href{http://arxiv.org/abs/1504.07551}{{\ttfamily arXiv:1504.07551
  [hep-ph]}}.

\bibitem{Arvanitaki:2016xds}
A.~Arvanitaki, S.~Dimopoulos, V.~Gorbenko, J.~Huang, and K.~Van~Tilburg, ``{A
  small weak scale from a small cosmological constant},''
  \href{http://dx.doi.org/10.1007/JHEP05(2017)071}{{\em JHEP} {\bfseries 05}
  (2017) 071}, \href{http://arxiv.org/abs/1609.06320}{{\ttfamily
  arXiv:1609.06320 [hep-ph]}}.

\bibitem{Banerjee:2018xmn}
A.~Banerjee, H.~Kim, and G.~Perez, ``{Coherent relaxion dark matter},''
  \href{http://dx.doi.org/10.1103/PhysRevD.100.115026}{{\em Phys. Rev. D}
  {\bfseries 100} no.~11, (2019) 115026},
  \href{http://arxiv.org/abs/1810.01889}{{\ttfamily arXiv:1810.01889
  [hep-ph]}}.

\bibitem{Banerjee:2020kww}
A.~Banerjee, H.~Kim, O.~Matsedonskyi, G.~Perez, and M.~S. Safronova, ``{Probing
  the Relaxed Relaxion at the Luminosity and Precision Frontiers},''
  \href{http://dx.doi.org/10.1007/JHEP07(2020)153}{{\em JHEP} {\bfseries 07}
  (2020) 153}, \href{http://arxiv.org/abs/2004.02899}{{\ttfamily
  arXiv:2004.02899 [hep-ph]}}.

\bibitem{Arkani-Hamed:2020yna}
N.~Arkani-Hamed, R.~T. D'Agnolo, and H.~D. Kim, ``{Weak scale as a trigger},''
  \href{http://dx.doi.org/10.1103/PhysRevD.104.095014}{{\em Phys. Rev. D}
  {\bfseries 104} no.~9, (2021) 095014},
  \href{http://arxiv.org/abs/2012.04652}{{\ttfamily arXiv:2012.04652
  [hep-ph]}}.

\bibitem{TitoDAgnolo:2021nhd}
R.~Tito~D'Agnolo and D.~Teresi, ``{Sliding Naturalness: New Solution to the
  Strong-$CP$ and Electroweak-Hierarchy Problems},''
  \href{http://dx.doi.org/10.1103/PhysRevLett.128.021803}{{\em Phys. Rev.
  Lett.} {\bfseries 128} no.~2, (2022) 021803},
  \href{http://arxiv.org/abs/2106.04591}{{\ttfamily arXiv:2106.04591
  [hep-ph]}}.

\bibitem{TitoDAgnolo:2021pjo}
R.~Tito~D'Agnolo and D.~Teresi, ``{Sliding naturalness: cosmological selection
  of the weak scale},'' \href{http://dx.doi.org/10.1007/JHEP02(2022)023}{{\em
  JHEP} {\bfseries 02} (2022) 023},
  \href{http://arxiv.org/abs/2109.13249}{{\ttfamily arXiv:2109.13249
  [hep-ph]}}.

\bibitem{Chatrchyan:2022pcb}
A.~Chatrchyan and G.~Servant, ``{The Stochastic Relaxion},''
  \href{http://arxiv.org/abs/2210.01148}{{\ttfamily arXiv:2210.01148
  [hep-ph]}}.

\bibitem{Chatrchyan:2022dpy}
A.~Chatrchyan and G.~Servant, ``{Relaxion Dark Matter from Stochastic
  Misalignment},'' \href{http://arxiv.org/abs/2211.15694}{{\ttfamily
  arXiv:2211.15694 [hep-ph]}}.

\bibitem{Svrcek:2006yi}
P.~Svrcek and E.~Witten, ``{Axions In String Theory},''
  \href{http://dx.doi.org/10.1088/1126-6708/2006/06/051}{{\em JHEP} {\bfseries
  06} (2006) 051}, \href{http://arxiv.org/abs/hep-th/0605206}{{\ttfamily
  arXiv:hep-th/0605206}}.

\bibitem{Arvanitaki:2009fg}
A.~Arvanitaki, S.~Dimopoulos, S.~Dubovsky, N.~Kaloper, and J.~March-Russell,
  ``{String Axiverse},''
  \href{http://dx.doi.org/10.1103/PhysRevD.81.123530}{{\em Phys. Rev. D}
  {\bfseries 81} (2010) 123530},
  \href{http://arxiv.org/abs/0905.4720}{{\ttfamily arXiv:0905.4720 [hep-th]}}.

\bibitem{Lee:2017qve}
J.-W. Lee, ``{Brief History of Ultra-light Scalar Dark Matter Models},''
  \href{http://dx.doi.org/10.1051/epjconf/201816806005}{{\em EPJ Web Conf.}
  {\bfseries 168} (2018) 06005},
  \href{http://arxiv.org/abs/1704.05057}{{\ttfamily arXiv:1704.05057
  [astro-ph.CO]}}.

\bibitem{Hui:2021tkt}
L.~Hui, ``{Wave Dark Matter},''
  \href{http://dx.doi.org/10.1146/annurev-astro-120920-010024}{{\em Ann. Rev.
  Astron. Astrophys.} {\bfseries 59} (2021) 247--289},
  \href{http://arxiv.org/abs/2101.11735}{{\ttfamily arXiv:2101.11735
  [astro-ph.CO]}}.

\bibitem{2008gady.book.....B}
J.~{Binney} and S.~{Tremaine}, {\em {Galactic Dynamics: Second Edition}}.
\newblock 2008.

\bibitem{Merritt:2003qc}
D.~Merritt, ``{Single and binary black holes and their influence on nuclear
  structure},'' in {\em {Carnegie Observatories Centennial Symposium. 1.
  Coevolution of Black Holes and Galaxies}}.
\newblock 1, 2003.
\newblock \href{http://arxiv.org/abs/astro-ph/0301257}{{\ttfamily
  arXiv:astro-ph/0301257}}.

\bibitem{Sadeghian:2013laa}
L.~Sadeghian, F.~Ferrer, and C.~M. Will, ``{Dark matter distributions around
  massive black holes: A general relativistic analysis},''
  \href{http://dx.doi.org/10.1103/PhysRevD.88.063522}{{\em Phys. Rev. D}
  {\bfseries 88} no.~6, (2013) 063522},
  \href{http://arxiv.org/abs/1305.2619}{{\ttfamily arXiv:1305.2619
  [astro-ph.GA]}}.

\bibitem{Schive:2014hza}
H.-Y. Schive, M.-H. Liao, T.-P. Woo, S.-K. Wong, T.~Chiueh, T.~Broadhurst, and
  W.~Y.~P. Hwang, ``{Understanding the Core-Halo Relation of Quantum Wave Dark
  Matter from 3D Simulations},''
  \href{http://dx.doi.org/10.1103/PhysRevLett.113.261302}{{\em Phys. Rev.
  Lett.} {\bfseries 113} no.~26, (2014) 261302},
  \href{http://arxiv.org/abs/1407.7762}{{\ttfamily arXiv:1407.7762
  [astro-ph.GA]}}.

\bibitem{Chavanis:2011zi}
P.-H. Chavanis, ``{Mass-radius relation of Newtonian self-gravitating
  Bose-Einstein condensates with short-range interactions: I. Analytical
  results},'' \href{http://dx.doi.org/10.1103/PhysRevD.84.043531}{{\em Phys.
  Rev. D} {\bfseries 84} (2011) 043531},
  \href{http://arxiv.org/abs/1103.2050}{{\ttfamily arXiv:1103.2050
  [astro-ph.CO]}}.

\bibitem{Hui:2016ltb}
L.~Hui, J.~P. Ostriker, S.~Tremaine, and E.~Witten, ``{Ultralight scalars as
  cosmological dark matter},''
  \href{http://dx.doi.org/10.1103/PhysRevD.95.043541}{{\em Phys. Rev. D}
  {\bfseries 95} no.~4, (2017) 043541},
  \href{http://arxiv.org/abs/1610.08297}{{\ttfamily arXiv:1610.08297
  [astro-ph.CO]}}.

\bibitem{Schive:2014dra}
H.-Y. Schive, T.~Chiueh, and T.~Broadhurst, ``{Cosmic Structure as the Quantum
  Interference of a Coherent Dark Wave},''
  \href{http://dx.doi.org/10.1038/nphys2996}{{\em Nature Phys.} {\bfseries 10}
  (2014) 496--499}, \href{http://arxiv.org/abs/1406.6586}{{\ttfamily
  arXiv:1406.6586 [astro-ph.GA]}}.

\bibitem{Bar:2018acw}
N.~Bar, D.~Blas, K.~Blum, and S.~Sibiryakov, ``{Galactic rotation curves versus
  ultralight dark matter: Implications of the soliton-host halo relation},''
  \href{http://dx.doi.org/10.1103/PhysRevD.98.083027}{{\em Phys. Rev. D}
  {\bfseries 98} no.~8, (2018) 083027},
  \href{http://arxiv.org/abs/1805.00122}{{\ttfamily arXiv:1805.00122
  [astro-ph.CO]}}.

\bibitem{Bar:2019bqz}
N.~Bar, K.~Blum, J.~Eby, and R.~Sato, ``{Ultralight dark matter in disk
  galaxies},'' \href{http://dx.doi.org/10.1103/PhysRevD.99.103020}{{\em Phys.
  Rev. D} {\bfseries 99} no.~10, (2019) 103020},
  \href{http://arxiv.org/abs/1903.03402}{{\ttfamily arXiv:1903.03402
  [astro-ph.CO]}}.

\bibitem{Bar:2021kti}
N.~Bar, K.~Blum, and C.~Sun, ``{Galactic rotation curves versus ultralight dark
  matter: A systematic comparison with SPARC data},''
  \href{http://dx.doi.org/10.1103/PhysRevD.105.083015}{{\em Phys. Rev. D}
  {\bfseries 105} no.~8, (2022) 083015},
  \href{http://arxiv.org/abs/2111.03070}{{\ttfamily arXiv:2111.03070
  [hep-ph]}}.

\bibitem{Chan:2021bja}
H.~Y.~J. Chan, E.~G.~M. Ferreira, S.~May, K.~Hayashi, and M.~Chiba, ``{The
  diversity of core\textendash{}halo structure in the fuzzy dark matter
  model},'' \href{http://dx.doi.org/10.1093/mnras/stac063}{{\em Mon. Not. Roy.
  Astron. Soc.} {\bfseries 511} no.~1, (2022) 943--952},
  \href{http://arxiv.org/abs/2110.11882}{{\ttfamily arXiv:2110.11882
  [astro-ph.CO]}}.

\bibitem{Bar:2019pnz}
N.~Bar, K.~Blum, T.~Lacroix, and P.~Panci, ``{Looking for ultralight dark
  matter near supermassive black holes},''
  \href{http://dx.doi.org/10.1088/1475-7516/2019/07/045}{{\em JCAP} {\bfseries
  07} (2019) 045}, \href{http://arxiv.org/abs/1905.11745}{{\ttfamily
  arXiv:1905.11745 [astro-ph.CO]}}.

\bibitem{Mocz:2017wlg}
P.~Mocz, M.~Vogelsberger, V.~H. Robles, J.~Zavala, M.~Boylan-Kolchin,
  A.~Fialkov, and L.~Hernquist, ``{Galaxy formation with BECDM \textendash{} I.
  Turbulence and relaxation of idealized haloes},''
  \href{http://dx.doi.org/10.1093/mnras/stx1887}{{\em Mon. Not. Roy. Astron.
  Soc.} {\bfseries 471} no.~4, (2017) 4559--4570},
  \href{http://arxiv.org/abs/1705.05845}{{\ttfamily arXiv:1705.05845
  [astro-ph.CO]}}.

\bibitem{Veltmaat:2018dfz}
J.~Veltmaat, J.~C. Niemeyer, and B.~Schwabe, ``{Formation and structure of
  ultralight bosonic dark matter halos},''
  \href{http://dx.doi.org/10.1103/PhysRevD.98.043509}{{\em Phys. Rev. D}
  {\bfseries 98} no.~4, (2018) 043509},
  \href{http://arxiv.org/abs/1804.09647}{{\ttfamily arXiv:1804.09647
  [astro-ph.CO]}}.

\bibitem{Lin:2018whl}
S.-C. Lin, H.-Y. Schive, S.-K. Wong, and T.~Chiueh, ``{Self-consistent
  construction of virialized wave dark matter halos},''
  \href{http://dx.doi.org/10.1103/PhysRevD.97.103523}{{\em Phys. Rev. D}
  {\bfseries 97} no.~10, (2018) 103523},
  \href{http://arxiv.org/abs/1801.02320}{{\ttfamily arXiv:1801.02320
  [astro-ph.CO]}}.

\bibitem{Yavetz:2021pbc}
T.~D. Yavetz, X.~Li, and L.~Hui, ``{Construction of wave dark matter halos:
  Numerical algorithm and analytical constraints},''
  \href{http://dx.doi.org/10.1103/PhysRevD.105.023512}{{\em Phys. Rev. D}
  {\bfseries 105} no.~2, (2022) 023512},
  \href{http://arxiv.org/abs/2109.06125}{{\ttfamily arXiv:2109.06125
  [astro-ph.CO]}}.

\bibitem{Kim:2021yyo}
H.~Kim and A.~Lenoci, ``{Gravitational focusing of wave dark matter},''
  \href{http://dx.doi.org/10.1103/PhysRevD.105.063032}{{\em Phys. Rev. D}
  {\bfseries 105} no.~6, (2022) 063032},
  \href{http://arxiv.org/abs/2112.05718}{{\ttfamily arXiv:2112.05718
  [hep-ph]}}.

\bibitem{Detweiler:1980uk}
S.~L. Detweiler, ``{KLEIN-GORDON EQUATION AND ROTATING BLACK HOLES},''
  \href{http://dx.doi.org/10.1103/PhysRevD.22.2323}{{\em Phys. Rev. D}
  {\bfseries 22} (1980) 2323--2326}.

\bibitem{Baumann:2019eav}
D.~Baumann, H.~S. Chia, J.~Stout, and L.~ter Haar, ``{The Spectra of
  Gravitational Atoms},''
  \href{http://dx.doi.org/10.1088/1475-7516/2019/12/006}{{\em JCAP} {\bfseries
  12} (2019) 006}, \href{http://arxiv.org/abs/1908.10370}{{\ttfamily
  arXiv:1908.10370 [gr-qc]}}.

\bibitem{Cardoso:2022nzc}
V.~Cardoso, T.~Ikeda, R.~Vicente, and M.~Zilh\~ao, ``{Parasitic black holes:
  The swallowing of a fuzzy dark matter soliton},''
  \href{http://dx.doi.org/10.1103/PhysRevD.106.L121302}{{\em Phys. Rev. D}
  {\bfseries 106} no.~12, (2022) L121302},
  \href{http://arxiv.org/abs/2207.09469}{{\ttfamily arXiv:2207.09469 [gr-qc]}}.

\bibitem{Madau:2001sc}
P.~Madau and M.~J. Rees, ``{Massive black holes as Population III remnants},''
  \href{http://dx.doi.org/10.1086/319848}{{\em Astrophys. J. Lett.} {\bfseries
  551} (2001) L27--L30},
  \href{http://arxiv.org/abs/astro-ph/0101223}{{\ttfamily
  arXiv:astro-ph/0101223}}.

\bibitem{Koushiappas:2003zn}
S.~M. Koushiappas, J.~S. Bullock, and A.~Dekel, ``{Massive black hole seeds
  from low angular momentum material},''
  \href{http://dx.doi.org/10.1111/j.1365-2966.2004.08190.x}{{\em Mon. Not. Roy.
  Astron. Soc.} {\bfseries 354} (2004) 292},
  \href{http://arxiv.org/abs/astro-ph/0311487}{{\ttfamily
  arXiv:astro-ph/0311487}}.

\bibitem{Greene:2019vlv}
J.~E. Greene, J.~Strader, and L.~C. Ho, ``{Intermediate-Mass Black Holes},''
  \href{http://dx.doi.org/10.1146/annurev-astro-032620-021835}{{\em Ann. Rev.
  Astron. Astrophys.} {\bfseries 58} (2020) 257--312},
  \href{http://arxiv.org/abs/1911.09678}{{\ttfamily arXiv:1911.09678
  [astro-ph.GA]}}.

\bibitem{Ullio:2001fb}
P.~Ullio, H.~Zhao, and M.~Kamionkowski, ``{A Dark matter spike at the galactic
  center?},'' \href{http://dx.doi.org/10.1103/PhysRevD.64.043504}{{\em Phys.
  Rev. D} {\bfseries 64} (2001) 043504},
  \href{http://arxiv.org/abs/astro-ph/0101481}{{\ttfamily
  arXiv:astro-ph/0101481}}.

\bibitem{Merritt:2002vj}
D.~Merritt, M.~Milosavljevic, L.~Verde, and R.~Jimenez, ``{Dark matter spikes
  and annihilation radiation from the galactic center},''
  \href{http://dx.doi.org/10.1103/PhysRevLett.88.191301}{{\em Phys. Rev. Lett.}
  {\bfseries 88} (2002) 191301},
  \href{http://arxiv.org/abs/astro-ph/0201376}{{\ttfamily
  arXiv:astro-ph/0201376}}.

\bibitem{Merritt:2003qk}
D.~Merritt, ``{Evolution of the dark matter distribution at the galactic
  center},'' \href{http://dx.doi.org/10.1103/PhysRevLett.92.201304}{{\em Phys.
  Rev. Lett.} {\bfseries 92} (2004) 201304},
  \href{http://arxiv.org/abs/astro-ph/0311594}{{\ttfamily
  arXiv:astro-ph/0311594}}.

\bibitem{Bertone:2005xz}
G.~Bertone, A.~R. Zentner, and J.~Silk, ``{A new signature of dark matter
  annihilations: gamma-rays from intermediate-mass black holes},''
  \href{http://dx.doi.org/10.1103/PhysRevD.72.103517}{{\em Phys. Rev. D}
  {\bfseries 72} (2005) 103517},
  \href{http://arxiv.org/abs/astro-ph/0509565}{{\ttfamily
  arXiv:astro-ph/0509565}}.

\bibitem{Bertone:2009kj}
G.~Bertone, M.~Fornasa, M.~Taoso, and A.~Zentner, ``{Dark Matter Annihilation
  around Intermediate Mass Black Holes: an update},''
  \href{http://dx.doi.org/10.1088/1367-2630/11/10/105016}{{\em New J. Phys.}
  {\bfseries 11} (2009) 105016},
  \href{http://arxiv.org/abs/0905.4736}{{\ttfamily arXiv:0905.4736
  [astro-ph.HE]}}.

\bibitem{Fan:2006dp}
X.-H. Fan, C.~L. Carilli, and B.~G. Keating, ``{Observational constraints on
  cosmic reionization},''
  \href{http://dx.doi.org/10.1146/annurev.astro.44.051905.092514}{{\em Ann.
  Rev. Astron. Astrophys.} {\bfseries 44} (2006) 415--462},
  \href{http://arxiv.org/abs/astro-ph/0602375}{{\ttfamily
  arXiv:astro-ph/0602375}}.

\bibitem{Mortlock:2011va}
D.~J. Mortlock {\em et~al.}, ``{A luminous quasar at a redshift of z =
  7.085},'' \href{http://dx.doi.org/10.1038/nature10159}{{\em Nature}
  {\bfseries 474} (2011) 616}, \href{http://arxiv.org/abs/1106.6088}{{\ttfamily
  arXiv:1106.6088 [astro-ph.CO]}}.

\bibitem{Banados:2017unc}
E.~Banados {\em et~al.}, ``{An 800-million-solar-mass black hole in a
  significantly neutral Universe at redshift 7.5}''
  \href{http://dx.doi.org/10.1038/nature25180}{{\em Nature} {\bfseries 553}
  no.~7689, (2018) 473--476}, \href{http://arxiv.org/abs/1712.01860}{{\ttfamily
  arXiv:1712.01860 [astro-ph.GA]}}.

\bibitem{Becker:2021ivq}
N.~Becker, L.~Sagunski, L.~Prinz, and S.~Rastgoo, ``{Circularization versus
  eccentrification in intermediate mass ratio inspirals inside dark matter
  spikes},'' \href{http://dx.doi.org/10.1103/PhysRevD.105.063029}{{\em Phys.
  Rev. D} {\bfseries 105} no.~6, (2022) 063029},
  \href{http://arxiv.org/abs/2112.09586}{{\ttfamily arXiv:2112.09586 [gr-qc]}}.

\bibitem{Annulli:2020ilw}
L.~Annulli, V.~Cardoso, and R.~Vicente, ``{Stirred and shaken: Dynamical
  behavior of boson stars and dark matter cores},''
  \href{http://dx.doi.org/10.1016/j.physletb.2020.135944}{{\em Phys. Lett. B}
  {\bfseries 811} (2020) 135944},
  \href{http://arxiv.org/abs/2007.03700}{{\ttfamily arXiv:2007.03700
  [astro-ph.HE]}}.

\bibitem{Annulli:2020lyc}
L.~Annulli, V.~Cardoso, and R.~Vicente, ``{Response of ultralight dark matter
  to supermassive black holes and binaries},''
  \href{http://dx.doi.org/10.1103/PhysRevD.102.063022}{{\em Phys. Rev. D}
  {\bfseries 102} no.~6, (2020) 063022},
  \href{http://arxiv.org/abs/2009.00012}{{\ttfamily arXiv:2009.00012 [gr-qc]}}.

\bibitem{Robson:2018ifk}
T.~Robson, N.~J. Cornish, and C.~Liu, ``{The construction and use of LISA
  sensitivity curves},'' \href{http://dx.doi.org/10.1088/1361-6382/ab1101}{{\em
  Class. Quant. Grav.} {\bfseries 36} no.~10, (2019) 105011},
  \href{http://arxiv.org/abs/1803.01944}{{\ttfamily arXiv:1803.01944
  [astro-ph.HE]}}.

\bibitem{Owen:1995tm}
B.~J. Owen, ``{Search templates for gravitational waves from inspiraling
  binaries: Choice of template spacing},''
  \href{http://dx.doi.org/10.1103/PhysRevD.53.6749}{{\em Phys. Rev. D}
  {\bfseries 53} (1996) 6749--6761},
  \href{http://arxiv.org/abs/gr-qc/9511032}{{\ttfamily arXiv:gr-qc/9511032}}.

\bibitem{Speagle:2019ivv}
J.~S. Speagle, ``{dynesty: a dynamic nested sampling package for estimating
  Bayesian posteriors and evidences},''
  \href{http://dx.doi.org/10.1093/mnras/staa278}{{\em Mon. Not. Roy. Astron.
  Soc.} {\bfseries 493} no.~3, (2020) 3132--3158},
  \href{http://arxiv.org/abs/1904.02180}{{\ttfamily arXiv:1904.02180
  [astro-ph.IM]}}.

\bibitem{Unruh:1976fm}
W.~G. Unruh, ``{Absorption Cross-Section of Small Black Holes},''
  \href{http://dx.doi.org/10.1103/PhysRevD.14.3251}{{\em Phys. Rev. D}
  {\bfseries 14} (1976) 3251--3259}.

\bibitem{Traykova:2021dua}
D.~Traykova, K.~Clough, T.~Helfer, E.~Berti, P.~G. Ferreira, and L.~Hui,
  ``{Dynamical friction from scalar dark matter in the relativistic regime},''
  \href{http://dx.doi.org/10.1103/PhysRevD.104.103014}{{\em Phys. Rev. D}
  {\bfseries 104} no.~10, (2021) 103014},
  \href{http://arxiv.org/abs/2106.08280}{{\ttfamily arXiv:2106.08280 [gr-qc]}}.

\bibitem{Vicente:2022ivh}
R.~Vicente and V.~Cardoso, ``{Dynamical friction of black holes in ultralight
  dark matter},'' \href{http://dx.doi.org/10.1103/PhysRevD.105.083008}{{\em
  Phys. Rev. D} {\bfseries 105} no.~8, (2022) 083008},
  \href{http://arxiv.org/abs/2201.08854}{{\ttfamily arXiv:2201.08854 [gr-qc]}}.

\bibitem{1987degc.book.....S}
L.~{Spitzer}, {\em {Dynamical evolution of globular clusters}}.
\newblock 1987.

\bibitem{Bar-Or:2018pxz}
B.~Bar-Or, J.-B. Fouvry, and S.~Tremaine, ``{Relaxation in a Fuzzy Dark Matter
  Halo},'' \href{http://dx.doi.org/10.3847/1538-4357/aaf28c}{{\em Astrophys.
  J.} {\bfseries 871} no.~1, (2019) 28},
  \href{http://arxiv.org/abs/1809.07673}{{\ttfamily arXiv:1809.07673
  [astro-ph.GA]}}.

\bibitem{Bar-Or:2020tys}
B.~Bar-Or, J.-B. Fouvry, and S.~Tremaine, ``{Relaxation in a Fuzzy Dark Matter
  Halo. II. Self-consistent Kinetic Equations},''
  \href{http://dx.doi.org/10.3847/1538-4357/abfb66}{{\em Astrophys. J.}
  {\bfseries 915} no.~1, (2021) 27},
  \href{http://arxiv.org/abs/2010.10212}{{\ttfamily arXiv:2010.10212
  [astro-ph.GA]}}.

\bibitem{Chavanis:2020upb}
P.-H. Chavanis, ``{Landau equation for self-gravitating classical and quantum
  particles: application to dark matter},''
  \href{http://dx.doi.org/10.1140/epjp/s13360-021-01617-3}{{\em Eur. Phys. J.
  Plus} {\bfseries 136} no.~6, (2021) 703},
  \href{http://arxiv.org/abs/2012.12858}{{\ttfamily arXiv:2012.12858
  [astro-ph.GA]}}.

\bibitem{Bar:2021jff}
N.~Bar, D.~Blas, K.~Blum, and H.~Kim, ``{Assessing the Fornax globular cluster
  timing problem in different models of dark matter},''
  \href{http://dx.doi.org/10.1103/PhysRevD.104.043021}{{\em Phys. Rev. D}
  {\bfseries 104} no.~4, (2021) 043021},
  \href{http://arxiv.org/abs/2102.11522}{{\ttfamily arXiv:2102.11522
  [astro-ph.GA]}}.

\end{thebibliography}\endgroup
\end{document}